# Landau-Ginzburg theory of charge density wave formation accompanying lattice and electronic long-range ordering

Anna N. Morozovska[1*], Eugene A. Eliseev[2], Venkatraman Gopalan[3†] and Long-Qing Chen[3‡]

[1] Institute of Physics, National Academy of Sciences of Ukraine,
41, pr. Nauki, 03028 Kyiv, Ukraine

[2] Institute for Problems of Materials Science, National Academy of Sciences of Ukraine,
3, Krjijanovskogo, 03142 Kyiv, Ukraine

[3] Department of Materials Science and Engineering,
Pennsylvania State University, University Park, PA 16802, USA

**Abstract**

We propose an analytical Landau-Ginzburg theory of the charge density waves coupled with lattice and electronic long-range order parameters. Examples of long-range order include electronic wave function of superconducting Cooper pairs, structural distortions, electric polarization, and magnetization. We formulate the Landau-Ginzburg free energy density as power expansion with respect to the charge density and other long-range order parameters, as well as their spatial gradients, and biquadratic coupling terms. We introduced a biquadratic coupling between the charge density gradient and long-range order parameters, as well as nonlinear higher gradients of the long-range order parameters. The biquadratic gradient coupling is critical to the appearance of different spatially-modulated phases in charge-ordered ferroics and high-temperature superconductors. We derived the thermodynamic conditions for the stability of the spatially-modulated phases, which are the intertwined spatial waves of charge density and lattice/electronic long-range order. The analytical expressions for the energies of different phases, corresponding order parameters, charge density waves amplitudes and modulation periods, obtained in this work, can be employed to guide the comprehensive physical explanation, deconvolution and Bayesian analysis of experimental data on quantum materials ranging from charge-ordered ferroics to high-temperature superconductors.

[*] anna.n.morozovska@gmail.com

[†] vgopalan@psu.edu

[‡] lqc3@psu.edu

# I. INTRODUCTION

The formation of charge density waves may lead to anomalies in the electro-physical properties of bulk and two-dimensional transition metal dichalcogenides [1, 2, 3], high-temperature superconducting cuprates [4, 5], resistive switching materials [6, 7], ferroics and multiferroics [8, 9] exhibiting long-range structural, magnetic [10], antiferrodistortive and/or polar [11, 12] orders, and electric charge ordering. In many cases, superconductivity, spin, structural, orbital or polar ordering, and charge density waves are competing orders [13], which can coexist and often become "coupled" or "intertwined" [14, 15, 16], manifesting a complex interplay arising from strong intra-order correlations. For instance, Beaud et al. [17] revealed that the relaxation of orbital ordering (pseudo-Jahn-Teller mode) and charge ordering are coupled in a perovskite manganite. Hamidian et al. [18] demonstrated the existence of spatial modulation of the Cooper pairs density in a superconductor naturally coupled with phonon modes.

Incommensurate charge density waves, which are periodic spatial modulations of the electronic density uncorrelated with the lattice period, are ubiquitous in multiferroics [11] and superconducting cuprates [5]. Kivelson group proposed an effective field theory of a layered system with incommensurate spin- and/or charge-density wave orders in hole-doped cuprates [19]. They also analyzed the interplay between a uniform superconducting and a pair-density-wave order parameter in the neighborhood of a vortex [20], and, using a phenomenological nonlinear sigma model, revealed that the intertwining of the two superconducting orders leads to a charge density modulation with the same periodicity as the pair-density-wave. Using the Landau-Ginzburg-Wilson theory of competing orders, Yu and Kivelson [21] demonstrated the generic occurrence of a "fragile" superconducting phase at low temperatures in the presence of weak charge-density-wave disorder, and proposed an explanation of the discovered "resilient" superconducting phase at high fields in underdoped $YBa_2Cu_3O_{6+x}$.

In this regard, ferroics and high-temperature superconductors in the presence of charge density waves are very sensitive to the spatial gradients of long-range order parameters [22, 23]. Among many theoretical approaches (see e.g. [24]), the Landau-Ginzburg (**LG**) type models [1, 2, 5, 13, 17, 18, 25], which account for the spatial gradients of order parameters, have played a central role in understanding the formation of incommensurate charge density waves, their dynamics and interaction with lattice and electronic long-range order parameters, such as (anti)ferromagnetic, (anti)ferroelectric, (anti)ferrodistortive and/or superconducting orders. In the general case, LG models that include implicit expressions for depolarization (demagnetization) fields are self-consistent. Also, LG models allow analytical descriptions of complex coupled problems, including phase diagrams and incommensurate modulation periods. Analytical LG models significantly simplify physical explanation. They can provide a comprehensive deconvolution of experimental data, are suitable for their Bayesian analysis [26, 27] and complex machine learning.

This work aims to propose an analytical LG theory of the coupled charge density waves and lattice/electronic long-range ordering in ferroics and high-temperature superconductors. We formulate the LG free energy power expansion with respect to the charge density and other long-range order parameter(s), e.g., a wave function of superconducting Cooper pairs, structural distortion, electric polarization, magnetization, their spatial gradients, and biquadratic coupling terms [1, 2, 5, 13]. The structure of gradient terms in the free energy functional, used in this work, is different from earlier LG models [1, 2, 5, 13]. In particular, we introduced a biquadratic coupling between the charge density gradient and lattice/electronic long-range order, and nonlinear higher gradients of the long-range orders. Our primary interest are the determination of the appearance and stability conditions of the spatially-modulated phases [28], which are the coupled spatial waves of charge density and spontaneous (e.g., superconductive, polar or magnetic) long-range order. The gradient terms, introduced in this work, are critical to the appearance of the spatially-modulated phases.

We formulate the LG model in **Section II**. The phase diagrams, including the analytical expressions for stability conditions, amplitudes and modulation periods for the spatially-modulated phases, are presented in **Sections III** and **IV**. **Section V** provides the conclusions.

## II. PROBLEM FORMULATION

Following McMillan [1, 2] and Wandel et al [5], we consider complex order parameters to be vectorial and/or scalar, and use $\boldsymbol{\psi}_C$ and $\boldsymbol{\psi}_S$ to denote the charge density (**C**) and spontaneous (**S**) long-range orders, respectively. The primary purpose of this study is to analyze the spatially-modulated phases in a system described by the LG free energy functional of the $\boldsymbol{\psi}_C$ and $\boldsymbol{\psi}_S$ orders. These include the spatial waves of the charge density (**CW**), which is a $\boldsymbol{\psi}_C$ spatial modulation; spatial waves of the spontaneous long-range order (**SW**), which is a $\boldsymbol{\psi}_S$ spatial modulation; and intertwined spatial waves of charge density and spontaneous long-range order (**SCW**), which are coupled spatial modulation of both, $\boldsymbol{\psi}_C$ and $\boldsymbol{\psi}_S$. The total free energy of a system is then expressed as a functional of $\boldsymbol{\psi}_C$ and $\boldsymbol{\psi}_S$:

$$F[\boldsymbol{\psi}_C, \boldsymbol{\psi}_S] = \int (f_{int}[\boldsymbol{\psi}_C, \boldsymbol{\psi}_S] + f_{CO}[\boldsymbol{\psi}_C] + f_{SO}[\boldsymbol{\psi}_S]) d^3\vec{r}. \qquad (1a)$$

Following She et al [13], we assume that the interaction energy, $f_{int}[\boldsymbol{\psi}_C, \boldsymbol{\psi}_S]$, has the simplest form of biquadratic coupling of the order parameters, $f_{int} \sim \eta_{ij} |\psi_{Ci}|^2 |\psi_{Sj}|^2$, introduced by Haun [29], Salje et al [30], Balashova and Tagantsev [31]. Since the interaction between the order parameters gradients are very important, we also add the biquadratic gradient-coupling terms to $f_{int}[\boldsymbol{\psi}_C, \boldsymbol{\psi}_S]$, which acquires the form:

$$f_{int}[\boldsymbol{\psi}_C, \boldsymbol{\psi}_S] = \eta_{ij}|\psi_{Ci}|^2|\psi_{Sj}|^2 + \xi_{ijk}\left(|\psi_{Si}|^2\left|\frac{\partial \psi_{Cj}}{\partial x_k}\right|^2 + |\psi_{Ci}|^2\left|\frac{\partial \psi_{Sj}}{\partial x_k}\right|^2\right) + \chi_{ijkl}\left|\frac{\partial \psi_{Ci}}{\partial x_j}\right|^2 \left|\frac{\partial \psi_{Sk}}{\partial x_l}\right|^2. \qquad (1b)$$

Hereinafter the summation takes place over all repeated subscripts, $i$ $i, j, k = 1, 2$ or $3$ for vectorial order parameter(s).

Note that the flexo-type bilinear gradient-coupling terms, such as $\gamma_{ijk}\left(\psi_{Ci}\frac{\partial}{\partial x_j}\psi_{Sk} - \psi_{Sk}\frac{\partial}{\partial x_j}\psi_{Ci}\right)$, can exist for observable real quantities (see e.g., Ref. [11] and in Table I therein). The presence of a nonzero flexo-coupling tensor $\gamma_{ijk}$ depends strongly on the material spatial symmetry, tensorial and time-reversal properties of the order parameters [11]. Since the absolute value, $\left|\psi_{Ci}\frac{\partial}{\partial x_j}\psi_{Sk} - \psi_{Sk}\frac{\partial}{\partial x_j}\psi_{Ci}\right|$, is incompatible with minimization over $\boldsymbol{\psi}_C^*$ and $\boldsymbol{\psi}_S^*$, one should consider the complex form of the flexo-type bilinear gradient-coupling, e.g., $\frac{\gamma_{ijk}}{2}\left[\left(\psi_{Ci}^*\frac{\partial}{\partial x_j}\psi_{Sk} - \psi_{Sk}\frac{\partial}{\partial x_j}\psi_{Ci}^*\right) + \left(\psi_{Ci}\frac{\partial}{\partial x_j}\psi_{Sk}^* - \psi_{Sk}^*\frac{\partial}{\partial x_j}\psi_{Ci}\right)\right]$, where the tensor $\gamma_{ijk}$ is very sensitive to the symmetry, tensorial and time-reversal properties of $\boldsymbol{\psi}_C$ and $\boldsymbol{\psi}_S$. For instance, in a one-dimensional approximation for scalar functions $\psi_C$ and $\psi_S$, the term $\frac{\gamma}{2}\left[\left(\psi_C^*\frac{\partial}{\partial x}\psi_S - \psi_S\frac{\partial}{\partial x}\psi_C^*\right) + \left(\psi_C\frac{\partial}{\partial x}\psi_S^* - \psi_S^*\frac{\partial}{\partial x}\psi_C\right)\right]$ changes it sign under the x-inversion operation, $x \to -x$. Since the free energy of the parent (disordered) phase should be invariant with respect to the spatial inversion, the flexo-coupling coefficient $\gamma = 0$. The coefficient $\gamma$ can be nonzero when e.g., $\psi_C$ is a scalar and $\psi_S$ is an x-component of a polar vector; or $\psi_C$ is a scalar and $\psi_S$ is a pseudo-scalar; or $\psi_C$ is an x-component of an axial vector and $\psi_S$ is an x-component of a polar vector. These specific cases will be considered elsewhere, below we consider only the biquadratic gradient-coupling terms in $f_{int}$, which are nonzero for arbitrary symmetry, tensorial and time-reversal properties of $\boldsymbol{\psi}_C$ and $\boldsymbol{\psi}_S$.

As suggested by She et al [13], the contributions $f_{CO}$ and $f_{SO}$ have the following form:

$$f_{CO}[\boldsymbol{\psi}_C] = a_{Ci}|\psi_{Ci}|^2 + b_{Cij}|\psi_{Ci}|^2|\psi_{Cj}|^2 + g_{Cij}\left|\frac{\partial\psi_{Ci}}{\partial x_j}\right|^2 + \left(w_{Cij}\left|\frac{\partial\psi_{Ci}}{\partial x_j}\right|^2 + v_{Cij}\left|\frac{\partial^2\psi_{Ci}}{\partial x_j^2}\right|^2\right)|\psi_{Ci}|^2, \quad (1c)$$

$$f_{SO}[\boldsymbol{\psi}_S] = a_{Si}|\psi_{Si}|^2 + b_{Sij}|\psi_{Si}|^2|\psi_{Sj}|^2 + g_{Sij}\left|\frac{\partial\psi_{Si}}{\partial x_j}\right|^2 + \left(w_{Sij}\left|\frac{\partial\psi_{Si}}{\partial x_j}\right|^2 + v_{Sij}\left|\frac{\partial^2\psi_{Si}}{\partial x_j^2}\right|^2\right)|\psi_{Si}|^2. \quad (1d)$$

Due to the presence of biquadratic gradient-coupling terms in Eq.(1b), we need to add the higher gradient terms in Eq.(1c)-(1d) too. Note, that the structure of gradient terms in Eqs.(1b)-(1d) is a principal difference of this work in comparison with McMillan [1, 2], Wandel et al [5], and She et al [13]. In particular, they did not consider gradient-coupling terms and higher gradients, both of which, as it will be shown below, can have critical influence on the appearance and stability of separated and coupled spatially-modulated phases.

As usual in the Landau approach, the coefficients $a_{Si}$ and $a_{Ci}$ are assumed to depend linearly on the temperature $T$, and change their signs at critical temperatures, $T_S$ and $T_C$:

$$a_{Si}(T) = \alpha_{Si}\left(\frac{T}{T_S} - 1\right), \qquad a_{Ci}(T) = \alpha_{Ci}\left(\frac{T}{T_C} - 1\right), \qquad (2)$$

where the coefficients $\alpha_{Si} > 0$ and $\alpha_{Ci} > 0$.

For the thermodynamic stability of a system described by the functional (1), the matrix of coefficients $b_{Cij}$ and $b_{Sij}$ are positively defined; the matrices of higher gradient coefficients, $v_{Cij}$ and $v_{Sij}$, should be positively defined too, because the correlation energy of each subsystem, C and S, must be positive at high values of each order parameter.

The symmetry of vectors and tensors in Eqs.(1) is defined by the point group symmetry of the high temperature state of a material. Note, that, according to McMillan [1, 2], they can be coordinate-dependent to reflect the microstructure of the studied material. However, such a spatial dependence is contradictory to the space isotropy in the continuous Landau theory.

Following McMillan works [1, 2], the variation of a CW electron density $\rho$ can be introduced as:

$$\rho(\vec{r}, t) = \rho_0(1 + \text{Re}[\psi_{C1} + \psi_{C2} + \psi_{C3}]), \quad (3)$$

where the subscripts 1, 2 and 3 are the orthogonal crystal axes.

Note that Eq.(3) is specific to the CW, and it shows that an observable physical quantity, namely a positive electron density $\rho$, is proportional to the real part of a complex function $\boldsymbol{\psi}_C$. The relative variation of the electron density, $\delta\rho = \frac{\rho(\vec{r},t)}{\rho_0} - 1$, is equal to $\text{Re}[\psi_{C1} + \psi_{C2} + \psi_{C3}]$. Eq.(3) is similar to the concentration wave representation of atomic densities. At the same time, free energy functional (1) is much more general, since the complex order parameters, $\boldsymbol{\psi}_C$ and $\boldsymbol{\psi}_S$, can describe many physical variables, such as the vectorial concentration wave representation of atomic densities, the scalar charge density, the vector of spontaneous electric polarization in multiaxial ferroelectrics, or its component in uniaxial ferroelectrics, polar and antipolar order parameters in ferrielectrics and/or antiferroelectric, the axial vector of spontaneous magnetization and/or antiferromagnetic long-range order parameter(s) in ferromagnets, ferrimagnets and/or antiferromagnets, the axial vector(s) of aniferrodistortive long-range order in ferroic and multiferroic materials with spatially-modulated phases. We treated the long-range orders as complex variables, because this representation allows much easier and convenient way to find corresponding wave periods and amplitudes in the spatially-modulated SW, CW and SCW phases.

For the sake of simplicity, below we consider the one-component one-dimensional case: $\boldsymbol{\psi}_C(\vec{r}) \equiv \psi_C(x)$ and $\boldsymbol{\psi}_S(\vec{r}) \equiv \psi_S(x)$, which allow to omit all subscripts in Eqs.(1)-(3) and significantly simplify the analysis of the thermodynamic stability conditions. In the one-component one-dimensional case, the free energy (1) is stable at high values of the order parameters under the conditions

$$b_C > 0, \quad b_S > 0, \quad \eta > -2\sqrt{b_S b_C}. \quad (4a)$$

Under these conditions, the free energy (1) can describe several spatially-homogeneous phases, namely the disordered (**D**) phase, spontaneous long-range (**S**) and charge density (**C**) orderings, their coexistence (**S/C**), and mixed (**SC**) state.

For a correct description of the spatially-modulated phases, **CW**, **SW** and **SCW**, the free energy (1) should be stable at high values of the order parameters gradients; which is possible if the parameters simultaneously satisfy the following conditions:

$$v_S > 0, \quad w_S > 0, \quad v_C > 0, \quad w_C > 0, \quad \chi > -2\sqrt{v_S v_C}, \quad \xi > -2\sqrt{w_S w_C}. \quad (4b)$$

The conditions

$$g_S > 0, \quad g_C > 0, \quad (4c)$$

make the appearance of the "decoupled" CW or SW waves less favorable in comparison with homogeneous C or S phases. The set of the stable phases that satisfy condition (4) are described in the **Table 1**.

Analytical description of the order parameter and charge density modulation in the SW-C, CW-S and SCW phases is possible within a harmonic approximation for the wave's spatial profile:

$$\psi_S = \delta\psi_{S0} + \delta\psi_S \exp(\pm ikx), \quad \psi_C = \delta\psi_{C0} + \delta\psi_C \exp(\pm iqx). \tag{5}$$

Here the expressions for the bases, $\delta\psi_{S0}$ and $\delta\psi_{C0}$, modulation amplitudes, $\delta\psi_S$ and $\delta\psi_C$, wavevectors, $k$ and $q$, and energies of the modulated phase follow from the minimization of the free energy (1), which acquires a simple form given by Eq.(A.1b) in **Appendix S1** [32]. In particular, the minimization conditions, $\frac{\delta F}{\delta q} = 0$ and $\frac{\delta F}{\delta k} = 0$, lead to the system of equations for $k$ and $q$, which, in addition to the trivial solution, $q = 0$ and/or $k = 0$, can have nontrivial solutions, $q \neq 0$ and/or $k \neq 0$.

Note that the waves phases are decoupled in the harmonic approximation (5); their coupling can appear when one account for anharmonicity. The decoupling leads to the virtual independence of the different modulation periods, $k$ and $q$. Also, we did not find "ripples" in this work, corresponding to the simultaneous validity of inequalities $\delta\psi_{S0} \neq 0$ and $\delta\psi_S \neq 0$ (and/or $\delta\psi_{C0} \neq 0$ and $\delta\psi_C \neq 0$) in Eq.(5). We conclude that such states exist for very narrow ranges of parameters, which do not allow analytical description, and are rarely observable in reality. Due to the absence of ripples, and in accordance with Eq.(3) and **Table 1**, the observable density of CW is proportional to $|\delta\psi_C|\cos(qx)$. The observable quantity corresponding to S-order is determined by its physical nature, e.g., the Cooper pairs density is proportional to the wave function density $|\psi_S|^2$; while a spatial modulation of a specific antiferrodistortion can be proportional to $|\delta\psi_S|\cos(kx)$.

### III. PHASE DIAGRAM OF THE CONSIDERED SYSTEM

Let us analyze the phase diagram. At first, we introduce the dimensionless order parameter amplitudes and wavenumbers in the free energy (1):

$$\varphi_C^2 = \frac{|\psi_C|^2}{\psi_{C0}^2}, \quad \varphi_S^2 = \frac{|\psi_S|^2}{\psi_{S0}^2}, \quad q_c = \sqrt{\frac{g_C}{\alpha_C}}q, \quad k_s = \sqrt{\frac{g_S}{\alpha_S}}k, \tag{6a}$$

where $\psi_{C0}^2 = \frac{\alpha_C}{2b_C}$ and $\psi_{S0}^2 = \frac{\alpha_S}{2b_S}$. Also, we introduce the following dimensionless parameters, gradient coefficients and biquadratic coupling constants:

$$f_c = \frac{\alpha_C^2}{b_C}, \quad f_s = \frac{\alpha_S^2}{b_S}, \quad \vartheta = \frac{f_s}{f_c}, \tag{6b}$$

$$v_c^* = \frac{v_C}{b_C}\left(\frac{\alpha_C}{g_C}\right)^2, \quad v_s^* = \frac{v_S}{b_S}\left(\frac{\alpha_S}{g_S}\right)^2, \quad w_c^* = \frac{w_C}{b_C}\frac{\alpha_C}{g_C}, \quad w_s^* = \frac{w_S}{b_S}\frac{\alpha_S}{g_S}, \tag{6c}$$

$$\eta^* = \frac{\alpha_C\alpha_S}{2b_Cb_S}\frac{\eta}{f_c}, \quad \xi_s^* = \frac{\alpha_C\alpha_S}{2b_Cb_S}\frac{\xi}{f_c}\frac{\alpha_S}{g_S}, \quad \xi_c^* = \frac{\alpha_C\alpha_S}{2b_Cb_S}\frac{\xi}{f_c}\frac{\alpha_C}{g_C}, \quad \chi^* = \chi\frac{(\alpha_C\alpha_S)^2}{2b_Cb_S}\frac{\xi}{f_c}\frac{1}{g_Cg_S}. \tag{6d}$$

Using the dimensionless variables and order parameters (6), we rewrite Eq.(1) as:

$$\frac{f}{f_c} = \left[\theta_C(T,q_c)\frac{\varphi_C^2}{2} + \beta_c(q_c)\frac{\varphi_C^4}{4}\right] + \vartheta\left[\theta_S(T,k_s)\frac{\varphi_S^2}{2} + \beta_s(k_s)\frac{\varphi_S^4}{4}\right] + \mu(k_s,q_c)\frac{\varphi_C^2\varphi_S^2}{2}, \tag{7}$$

where we introduced the temperature- and wavenumber-dependent dimensionless functions:

$$\theta_C(T,q_c) = \frac{T}{T_C} - 1 + q_c^2, \qquad \theta_S(T,k_s) = \frac{T}{T_S} - 1 + k_s^2, \tag{8a}$$

$$\beta_c(q_c) = 1 + w_c^* q_c^2 + v_c^* q_c^4, \qquad \beta_s(k_s) = 1 + w_s^* k_s^2 + v_s^* k_s^4, \qquad (8b)$$

$$\mu(k_s, q_c) = \eta^* + \xi_s^* k_s^2 + \xi_c^* q_c^2 + \chi^* k_s^2 q_c^2. \qquad (8c)$$

The free energy (7) depends on the S to C subsystem energy ratio $\vartheta$, whose magnitude can be arbitrary: small, close to unity, or large; and these cases are analyzed below. The functions $\theta_C(T, q_c)$ and $\theta_S(T, k_s)$ are the reduced temperatures of C and S subsystems renormalized by the gradient energy of each order parameter. Positive functions $\beta_c(q_c)$ and $\beta_s(k_s)$ are fourth order nonlinearity of C and S subsystems renormalized by the gradient energy of each order parameter. The function $\mu(k_s, q_c)$ is the biquadratic coupling of S and C subsystems renormalized by the gradient coupling parameters $\xi_s^*$, $\xi_c^*$, and $\chi^*$, and wavenumbers, $k_s^2$ and $q_c^2$.

The free energy (7) is stable at high values of the order parameters and wavenumbers if the following conditions are simultaneously satisfied,

$$\eta^* > -\sqrt{\vartheta}, \ \xi_c^* > -\sqrt{v_c^*}, \ \xi_s^* > -\sqrt{\vartheta v_s^*}, \ \chi^* > -\sqrt{\vartheta v_c^* v_s^*}, \ w_c^* > -\sqrt{4v_s^*}, \ w_c^* > -\sqrt{4v_c^*}. \qquad (9)$$

Since the conditions (4) are satisfied, the last three inequalities are valid too.

The free energy (7) allows us to derive analytical expressions for the energies of different phases, the amplitudes of the corresponding order parameters and charge density waves, and modulation periods. Thermodynamically stable homogeneous phases and spatially-modulated states of the energy (7), corresponding absolute values of the order parameters, phase energies and stability conditions are listed in **Table 1.**

**Table 1.** Possible thermodynamically stable phases of the free energy (7)

| Phase/order | Order parameters values or/and amplitudes | Free energy density | Stability conditions |
|---|---|---|---|
| Disordered (**D**) | $\varphi_C = \varphi_S = 0$ | 0 | $\theta_C > 0, \theta_S > 0$ |
| Homogeneous spontaneous order (**S**) | $\varphi_S = \pm\sqrt{-\theta_S},$ $\varphi_C = 0$ | $f_{SO} = -\vartheta \frac{\theta_S^2}{4}$ | $f_{SO} = min,$ $\theta_S < 0$ |
| Homogeneous charge ordering (**C**) | $\varphi_C = \pm\sqrt{-\theta_C},$ $\varphi_S = 0.$ | $f_{CO} = -\frac{\theta_C^2}{4}$ | $f_{CO} = min,$ $\theta_C < 0$ |
| Mixed Homogeneous S and C orderings (**SC**) | $\varphi_C = \pm\sqrt{\vartheta \frac{\eta^*\theta_S - \theta_C}{\vartheta - \eta^{*2}}},$ $\varphi_S = \pm\sqrt{\frac{\eta^*\theta_C - \vartheta\theta_S}{\vartheta - \eta^{*2}}}.$ | $f_{SC} = -\vartheta \frac{\theta_C^2 - 2\eta^*\theta_C\theta_S + \vartheta\theta_S^2}{4(\vartheta - \eta^{*2})}$ | $f_{SC} = min,$ $\frac{\eta^*\theta_S - \theta_C}{\vartheta - \eta^{*2}} > 0,$ $\frac{\eta^*\theta_C - \vartheta\theta_S}{\vartheta - \eta^{*2}} > 0$ |
| Spontaneous (superconductive, polar or magnetic) long-range order waves (**SW**) | $\varphi_C = 0,$ $\varphi_S = \pm\sqrt{\frac{2(w_s^* - 2v_s^*\theta_S)}{4v_s^* - w_s^{*2}}},$ $k_s = \sqrt{\frac{-2 + w_s^*\theta_S}{w_s^* - 2v_s^*\theta_S}}.$ | $f_{SW} = \vartheta \frac{\theta_S(w_s^* - v_s^*\theta_S) - 1}{4v_s^* - w_s^{*2}},$ where $\theta_S = \frac{T}{T_S} - 1$ | $f_{SW} = min,$ $\frac{w_s^* - 2v_s^*\theta_S}{4v_s^* - w_s^{*2}} > 0,$ $\frac{-2 + w_s^*\theta_S}{w_s^* - 2v_s^*\theta_S} > 0.$ |
| Charge density waves (**CW**) | $\varphi_C = \pm\sqrt{2\frac{w_c^* - 2v_c^*\theta_C}{4v_c^* - w_c^{*2}}}, q_c = \sqrt{\frac{w_c^*\theta_C - 2}{w_c^* - 2v_c^*\theta_C}},$ $\varphi_S = 0.$ | $f_{CW} = \frac{\theta_C(w_c^* - v_c^*\theta_C) - 1}{4v_c^* - w_c^{*2}},$ where $\theta_C = \frac{T}{T_C} - 1$ | $f_{CW} = min,$ $\frac{w_c^* - 2v_c^*\theta_C}{4v_c^* - w_c^{*2}} > 0,$ $\frac{w_c^*\theta_C - 2}{w_c^* - 2v_c^*\theta_C} > 0.$ |

| Mixed charge density waves and homogeneous S-order (**CW-S**) | $\varphi_C = \pm\sqrt{\frac{4\vartheta v_C^*(\theta_C - \eta^*\theta_S) + 4\xi_S^*(\eta^* - \theta_C\xi_S^*) + 2\vartheta w_C^*(-1 + \theta_S\xi_S^*)}{4(\eta^{*2} - \vartheta)v_C^* + \vartheta w_C^{*2} - 4\eta^* w_C^*\xi_S^* + 4\xi_S^{*2}}}$, $q_c = \sqrt{\frac{-\vartheta w_C^*(\theta_C - \eta^*\theta_S) + 2(\vartheta - \eta^{*2} + (\eta^*\theta_C - \vartheta\theta_S)\xi_S^*)}{2\vartheta v_C^*(\theta_C - \eta^*\theta_S) + 2\xi_S^*(\eta^* - \theta_C\xi_S^*) + \vartheta w_C^*(\theta_S\xi_S^* - 1)}}$, $\varphi_S = \pm\sqrt{\frac{-\vartheta w_C^{*2}\theta_S - 4[v_C^*(\eta^*\theta_C - \vartheta\theta_S) + \xi_S^*] + 2w_C^*(\eta^* + \theta_C\xi_S^*)}{4(\eta^{*2} - \vartheta)v_C^* + \vartheta w_C^{*2} - 4\eta^* w_C^*\xi_S^* + 4\xi_S^{*2}}}$, $k_s = 0$ | Expression for $f_{CW-S}$ is given by Eq.(S.8) | $f_{CW-S} = min$ $q_c > 0$ $\varphi_C^2 > 0$ $\varphi_S^2 > 0$ |
|---|---|---|---|
| Mixed spontaneous long-range order waves and homogeneous C-order (**SW-C**) | $\varphi_C = \pm\sqrt{\vartheta\frac{-w_S^{*2}\theta_C + 4v_S^*(\theta_C - \eta^*\theta_S) - 4\xi_C^* + 2w_S^*(\eta^* + \theta_S\xi_C^*)}{4(\eta^{*2} - \vartheta)v_S^* + \vartheta w_S^{*2} - 4\eta^* w_S^*\xi_C^* + 4\xi_C^{*2}}}$, $q_c = 0$, $\varphi_S = \pm\sqrt{\frac{v_S^*(-4\eta^*\theta_C + 4\vartheta\theta_S) - 2w_S^*(\vartheta - \theta_C\xi_C^*) + 4\xi_C(\eta^* - \theta_S\xi_C^*)}{4(\eta^{*2} - \vartheta)v_S^* + \vartheta w_S^{*2} - 4\eta^* w_S^*\xi_C^* + 4\xi_C^{*2}}}$, $k_s = \sqrt{\frac{w_S^*(\eta^*\theta_C - \vartheta\theta_S) + 2[\vartheta - \eta^{*2} + (\eta^*\theta_S - \theta_C)\xi_C^*]}{2v_S^*(\vartheta\theta_S - \eta^*\theta_C) - w_S^*(\vartheta - \theta_C\xi_C^*) + 2\xi_C(\eta^* - \theta_S\xi_C^*)}}$ | Expression for $f_{SW-C}$ is given by Eq.(S.9) | $f_{SW-C} = min$ $k_s > 0$ $\varphi_C^2 > 0$ $\varphi_S^2 > 0$ |
| "Intertwined" or "coupled" waves of S and C orders (**SCW**) | $\varphi_C = \pm\sqrt{\frac{\Delta_C}{\Delta}}, \quad q_c = \sqrt{\frac{\Delta_q}{\Delta_C}},$ $\varphi_S = \pm\sqrt{\frac{\Delta_S}{\Delta}}, \quad k_s = \sqrt{\frac{\Delta_k}{\Delta_S}}.$ | Expressions for $f_{SCW}$, $\Delta, \Delta_C, \Delta_q, \Delta_S$, and $\Delta_k$ are given by Eqs.(S.10) | $f_{SCW} = min$, $\frac{\Delta_C}{\Delta} > 0, \frac{\Delta_S}{\Delta} > 0$, $\frac{\Delta_C}{\Delta} > 0, \frac{\Delta_S}{\Delta} > 0$. |

Using analytical expressions from **Table 1**, one can calculate the phase diagrams of various ferroics and high-Tc superconductors with charge order, lattice/electronic long-range order parameters, CW amplitudes and modulation periods. Note, that the relative density $\delta\rho$ of CW is proportional to $|\varphi_C|\cos(qx)$ in accordance with Eq.(3).

A concrete view of phase diagrams and related properties are determined by eleven dimensionless parameters: a constant of the S-C biquadratic coupling strength, $\eta^*$, three constants of the biquadratic gradient-coupling strength, $\xi_S^*, \xi_C^*$, and $\chi^*$, four gradient coefficients of S and C subsystems, $v_C^*, v_S^*, w_C^*$, and $w_S^*$, their energy ratio, $\vartheta$, and two temperature ratios, $\frac{T}{T_C}$ and $\frac{T}{T_S}$, respectively. Since we are mostly interested in the conditions of the spatially-modulated phases stability at corresponding equilibrium wavenumbers, we can take all possible measures to reduce the number of dimensionless parameters in the energy (7). For further reduction of the number of independent parameters up to eight, we analyzed the realistic case:

$$\xi_S^* = \xi_C^* = \xi^*, v_C^* = v_S^* = v^* \text{ and } w_C^* = w_S^* = w^*. \tag{10}$$

Putting $T \to 0$, we can exclude two transition temperatures, $T_S$ and $T_C$ from our consideration. As a result, the obtained six-parametric LG model is not much more complex than earlier LG models for two coupled long-range orders with respect to the number of fitting parameters.

The remained six parameters, $\eta^*, \xi^*, \vartheta, \chi^*, v^*$, and $w^*$, can provide a comprehensive deconvolution of versatile experimental data, as well as are suitable for their Bayesian analysis and machine learning. The phase diagram, order parameter amplitudes and modulation periods are most sensitive to the parameters, $\eta^*$ and $\xi^*$ (which can change their sign), and $\vartheta$, and less sensitive to the positive parameters $\chi^*$, $v^*$, and $w^*$. Several cases are presented and analyzed below.

Typical dependence of the free energy (7) on the order parameter amplitudes, $\varphi_C$ and $\varphi_S$, shown in **Figs. 1(a)-(d)**, is the appearance of the stable mixed SC phase at $\mu \leq 0$ independently on $\vartheta$ values. The coexistence of the S and C phases with relatively deep energy minima is possible only for $\vartheta = 1$ and $\mu > 1$; the stable S or C phase with deep energy minima appears at $\mu > 0$, and $\vartheta \gg 1$ or $\vartheta \ll 1$, respectively (see also **Figs. S1-S3** in **Supplement** [32]).

A typical phase diagram as a function of $\eta^*$ and $\vartheta$, calculated from the free energy Eq.(7) for $T = 0$, $\chi^* = 1$, $v^* = 10$, $w^* = 0.1$, and $\xi^* = -1$, is shown in **Figs. 1(e)**. Stable ordered phases are absent in the white region, where the inequality $\eta^* > -\sqrt{\vartheta}$ is invalid, and a system described by the free energy (7) is unstable. A relatively narrow stripe-shaped dark-green region of the SCW phase is located inside the largest light-green region of the SC state. Note that the area of the spatially-modulated CW-S, SW-C, and SCW phases for $\eta^* \leq 0$ decreases monotonically with decrease in $|\xi^*|$, and the phases almost disappear at $\xi^* \approx -0.20$ (see **Fig. S4** in **Appendix S3** [32]). The area of the homogeneous SC phase increases monotonically with $\xi^*$ increase from -1 to -0.25; and the phase occupies the regions of the modulated phases at more negative $\xi^*$. The location and area of the homogeneous S and C phase regions are independent of $\xi^*$. The S phase is stable at $\vartheta > 1$; the C phase is stable at $\vartheta < 1$, and the boundary between them is a horizonal line $\vartheta = 1$, as anticipated from **Table 1**.

Dimensionless wavenumbers, $q_c$ and $k_s$, corresponding to the diagram in **Fig. 1(e)**, are shown in **Figs. 1(g)-(h),** as a function of $\eta^*$ and $\vartheta$. The wavenumbers can be nonzero in the spatially-modulated phases, and are different, $q_c \neq k_s$, wherever $\vartheta \neq 1$. The wavenumber $q_c$ is nonzero in the SCW and CW-S phases, and $k_s$ is nonzero in the SCW and SW-C phases. The wavenumber $q_c$ tends to zero at the boundary between the CW-S and SC phases, and between the SCW and SW-C phases; it continuously changes across the CW-S and SCW border, increases strongly and reaches maxima at the boundary of the SCW phase with the unstable white region. The wavenumber $k_s$ tends to zero at the border between the SCW and CW-S phases; it continuously changes across the boundary between the SW-C and SCW phases, and increases strongly approaching the boundary between SW-C and SC phases.

Normalized order parameters (or their amplitudes in the spatially-modulated phases), $\varphi_c$ and $\varphi_s$, are shown in in **Figs. 1(k)-(l),** as a function of $\eta^*$ and $\vartheta$. As anticipated, the parameter $\varphi_c$ is zero in the S phase, and $\varphi_s$ is zero in the C phase. Both parameters are nonzero in the SCW, CW-S, SW-C, and SC phases, where their amplitudes are relatively small in comparison with the values near the boundary with the unstable region, where the order parameters formally diverge (see also **Fig. S5** in **Supplement** [32]).

A typical phase diagram, as a function of $\xi^*$ and $\vartheta$, calculated from the free energy Eq.(7) for $T = 0$, $\chi^* = 1$, $v^* = 10$, $w^* = 0.1$, and $\eta^* = -0.1$, is shown in **Fig. 1(f)**. Stable phases are absent in the white region, where the inequality $\xi^* > -\sqrt{\vartheta v^*}$ is invalid, and so a system described by the free energy (7) is unstable. The dark-green region of the SCW phase is largest for higher negative $\eta^*$; it decreases monotonically with increase in $\eta^*$ from negative values to zero; and, at the same time, the relatively small region of SW-C phase appears with $\eta^*$ increase from negative values to zero (see **Fig. S6** in **Supplement** [32]). The regions of the SC, CW-S and SW-C phases disappear at $\eta^* \geq 1$, being "adsorbed" by the regions of S and C phases. A very small triangular region of the SCW phase, located at the S-C boundary, $\vartheta = 1$, remains at $\eta^* = 1$ (see **Fig. S6(g)** in **Supplement** [32]).

Dimensionless wavenumbers, $q_c$ and $k_s$, corresponding to the diagram in **Fig. 1(f)**, are shown in **Fig. 1(i)-1(j),** as a function of $\xi^*$ and $\vartheta$. The wavenumber $q_c$ is nonzero in the SCW and CW-S phases, and $k_s$ is nonzero in the SCW and SW-C phases. The wavenumber $q_c$ tends to zero at the boundaries between the CW-S and SC phases, and/or between the SCW and SW-C phases; it continuously changes across the CW-S and SCW boundary, and increases approaching the boundary of the SCW phase with the unstable white region. The wavenumber $k_s$ tends to zero at the boundary between SCW and CW-S phases, and/or at the boundary between SW-C and SC phases; it continuously changes at the SCW and SW-C boundary, and increases strongly approaching the boundary between the SW-C phase and the white unstable region (see also **Fig. S4(c, f, i)** and **Fig. S6(c, f, i)** [32]).

Normalized order parameters (or/and their amplitudes in the spatially-modulated phases), $\varphi_c$ and $\varphi_s$, are shown in **Fig. 1(m)-1(n),** as a function of $\xi^*$ and $\vartheta$ (see also the middle and bottom rows of **Fig. S7** in **Supplement** [32]). As anticipated, the parameter $\varphi_c$ is zero in the S phase, and $\varphi_s$ is zero in the C phase. Both parameters are nonzero in the SCW, CW-S, SW-C, and SC phases, where their amplitudes are relatively small in comparison with the values near the boundary of SCW and/or SW-C phases with the unstable white region, where the order parameters formally diverge. Since the order parameters strongly increase when approaching the boundary with the unstable region, here the accuracy of harmonic approximation, used by us for the description of the long-range order in the spatially-modulated phases SCW, SW-C and CW-S, becomes very low and this questions on its applicability. Therefore, a very narrow region of the SW-C phase, which exists between the SCW phase and the white unstable region in **Fig. 1(e)** and **1(f)**, being not a numerical error, is rather an artifact related with the overestimation of model applicability limits.

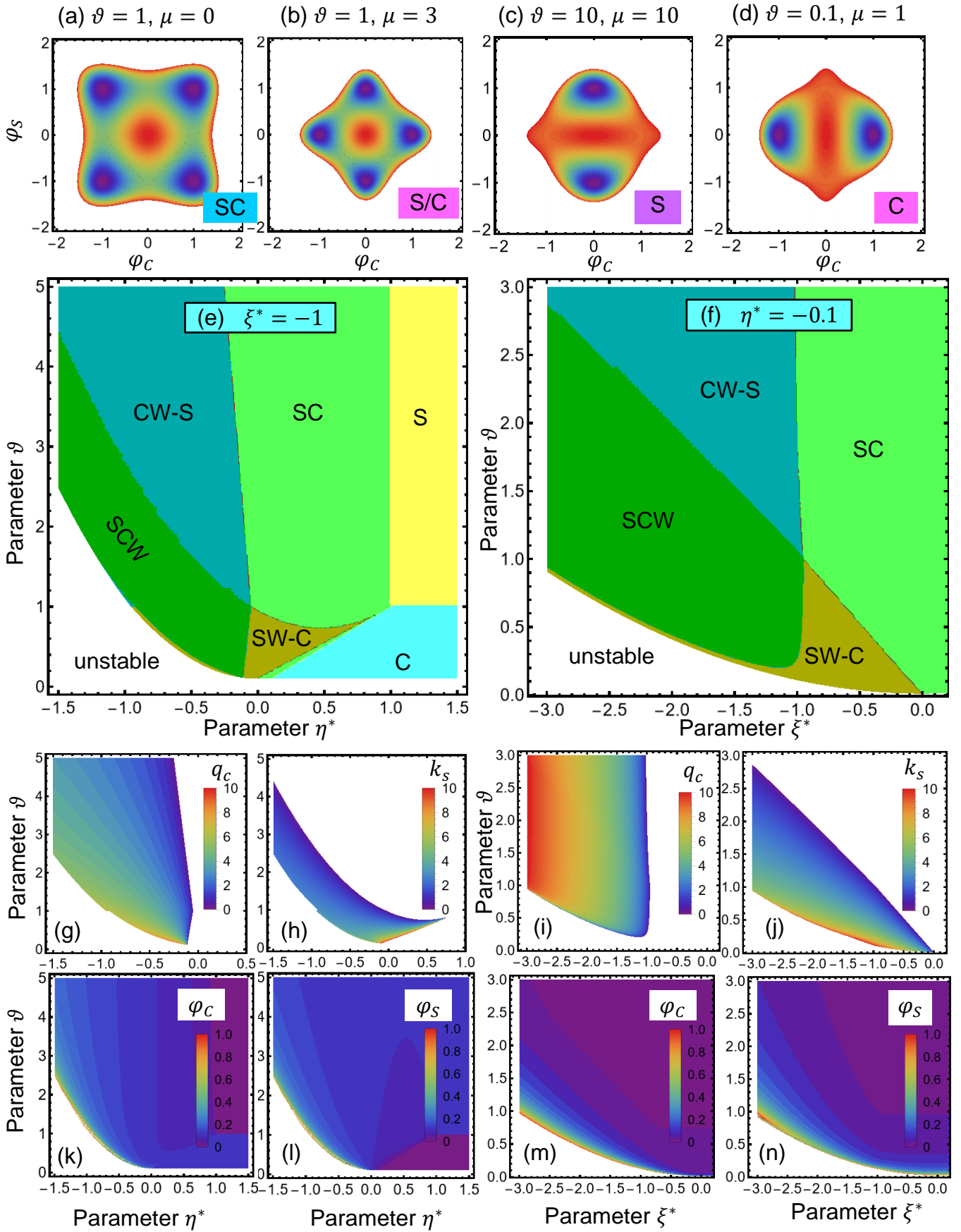

**FIGURE 1.** (a) − (d) The free energy (7), as a function of order parameter amplitudes, $\varphi_C$ and $\varphi_S$, calculated for different values of dimensionless parameters $\vartheta$ and $\mu$ listed near the plots. Red color denotes zero energy, while violet color is its minimal density in relative units. Phase diagrams (e, f), corresponding dimensionless wavenumbers, $q_c$ (g, i) and $k_s$ (h, j), and normalized order parameters (or/and their amplitudes in the modulated phases), $\varphi_C$ (k, m) and $\varphi_S$

(**l, n**), as a function of coupling constants, $\xi^*$ or $\eta^*$, and $\vartheta$, calculated for $T = 0$, $\chi^* = 1$, $v^* = 10$, $w^* = 0.1$, $\xi^* = -1$ for plots (**e, g, h**), and $\eta^* = -0.1$ for plots (**f, i, j**). Capital letters in the plots (a)-(d), (e) and (f) denote the mixed spontaneous long-range order - charge order state (SC), their coexistence (S/C), spontaneous long-range order (S) and its waves (SW), charge ordered (C) states and charge density waves (CW), and intertwined long-range order - charge density waves (SCW), respectively.

Results, presented above, correspond to very low temperatures, $0 \leq T \ll \min[T_C, T_S]$. The values of $T_C$ and/or $T_S$ can play a very significant role when the temperature $T$ rises, and becomes comparable with $\min[T_C, T_S]$. We hope to consider this question in details in near future. Preliminary calculations for higher temperatures, some results of which are presented in **Figs. A1** and **A2** in **Appendix A**, show that the ratio $T_C/T_S$ has a principal influence on the form of the phase diagrams. The temperature dependence of phases is very sensitive to the parameters, $\eta^*$ and $\vartheta$, relatively sensitive to the parameter $\xi^*$, and almost insensitive to the positive parameter $\chi^*$ for given values of $v^*$ and $w^*$.

## IV. NONLINEAR CHARGE DENSITY WAVES AND COMPLEX TOPOLOGICAL STRUCTURES

Note that the phase boundaries of the spatially-modulated phases, shown in **Fig. 1** and **Figs. S4-S7** in **Appendix S3** [32], are calculated within the harmonic approximation (5), which is valid near the boundaries of the second order phase transitions, where the CW profiles are sinusoidal or "soft". When one moves "deeper" into the spatially-modulated phase region, the harmonic approximation becomes invalid and the wave profiles become anharmonic. In a simplest 1D case a sinusoidal wave profile becomes "harder" and transforms into nonlinear elliptic functions, e.g., in an elliptic sine ("snoid"). In a more complex 2D and 3D cases, versatile topological structures (e.g., vortices, merons, skyrmions, and labyrinths) and other topological defects (e.g., ribbons, random spots, bubbles) can appear spontaneously near impurity atoms and/or vacancies, surfaces and/or interfaces. They are long-living metastable configurations, which minimize the system electrostatic energy. Within the framework of the proposed LG model, the appearance of the topological features are controlled by the structure and magnitude of the gradient terms in Eqs.(1b)-(1d); often being a gradient-induced morphological phase transition [33].

Let us underline that the structure of gradient terms in Eqs.(1b)-(1d) is more complete in comparison with earlier Landau-type models [1, 2, 13], because corresponding gradient terms in expressions (1b) and (1c) include the same, as well as additional terms. Due to the additions, which have critical influence on the appearance and stability of separated and coupled spatially-modulated phases, the proposed LG model describes more scenarios of the CW coexistence, competition or spatial separation with the other long-range order S in comparison with the models [1, 2, 13], and so it can potentially describe more experimental data.

Wandel et. al. [5] consider a different gradient term, $\frac{1}{2mQ^2}|\mathbf{Q} \cdot (\nabla\alpha - i\mathbf{Q})|^2$, which include the gradient of the real dimensionless order parameter $\alpha$, and the wave vector $\mathbf{Q}$ related to the wavelength of

incommensurate CW. The expression may be a particular case of more general expressions $g_{Cij}\left|\frac{\partial \psi_{Ci}}{\partial x_j}\right|^2$ and $g_{Sij}\left|\frac{\partial \psi_{Si}}{\partial x_j}\right|^2$ included in our Eq.(1c) and (1d) [34]. Wandel et. al [5] observe the enhancement of the CW spatial coherence in the high-Tc superconducting cuprate $YBa_2Cu_3O_{6+x}$ (YBCO) triggered by the laser-driven quench of the superconducting state and discussed three possible scenarios of the superconductivity-CW interaction. As the first scenario, they consider time-dependent LG model to interpret the dynamical interplay between interacting orders, assuming locally coexisting orders. In their formulation, which predicts homogenous and competitive superconductive and charge orders, the CW order parameter amplitude would increase on picosecond timescale, driven by the quench of superconductivity. The result was in principal disagreement with their observations, where the signal was dominated by a change of correlation length. They conclude that a simple competition model is incompatible with experimental results. Their second scenario considers phase-separated strongly competitive orders using modified McMillan [1, 2] formalism. In this case well-separated CW and superconducting domains, with an average spacing between neighboring CW domains larger than the CW periodicity, also remains incompatible with the experimental data. The third hypothesis, which well-explained their observations, is a superconductive region acting as a topological defect (e.g., dislocation) inside the CW region before photoexcitation. The topological defect induces a phase shift of the CW pattern, which propagates from the core of the defect, until the sudden photo-quench of superconductivity removes the defect.

Within harmonic approximation, the proposed LG model describes coexisting CW and S order, abbreviated as the CW-S phase and shown schematically in **Fig.2(a)**. The spatially-modulated S and C waves, which phase and periods are decoupled in the harmonic approximation are shown schematically in **Fig.2(b).** The spatially-separated S and C domains, which may look as small S phase regions inside the CW phase (or vice versa), can appear at the morphotropic boundary between C and S phases [shown schematically in **Fig.2(c)**]. All these scenarios are possible at $\eta^* < 0$, $\xi^* < 0$ and $\vartheta > 1$.

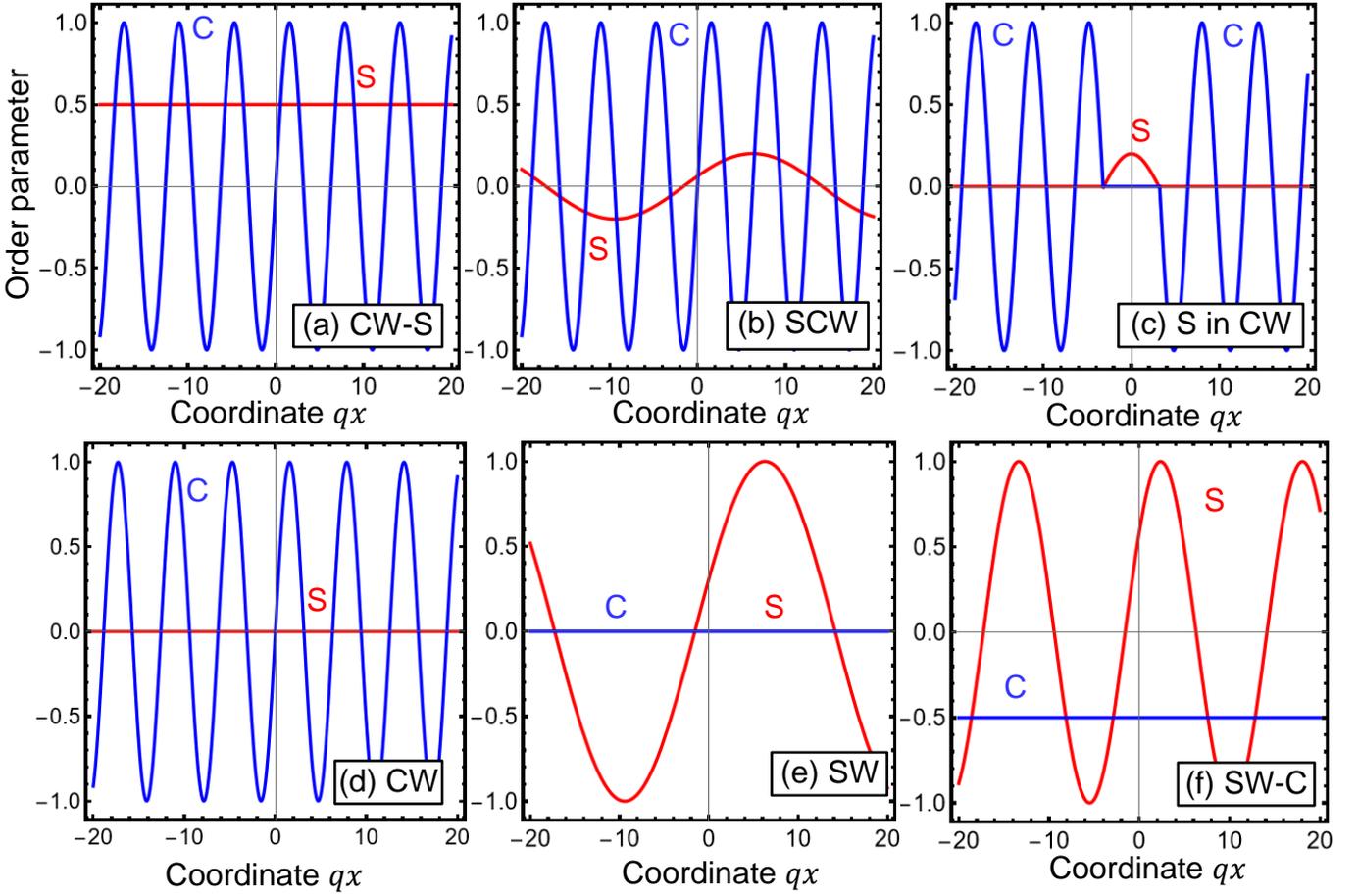

**FIGURE 2.** Schematical illustrations of **(a)** coexisting homogeneous S and CW orders in the CW-S phase; **(b)** the spatially-modulated S and C waves in the SCW phase; **(c)** the S phase region inside the CW phase. The spatially-modulated CW, SW and SW-C phases are shown in plots **(d)**, **(e)** and **(f)**, respectively. Red and blue curves are $Re[\psi_S]$ and $Re[\psi_C]$, respectively.

Anharmonicity and strong nonlinearity can change the situation, shown in **Fig.2(b)**-**2(c),** and the low-dimensional spatially-separated S domains can behave as 1D, 2D or 3D topological defects for CWs. To study the equilibrium states of spatially inhomogeneous structures, such as the topological defects, in materials with coupled long-range orders numerical modelling based on the minimization of the free energy (1) is required. Allowing for the Khalatnikov mechanism of the order parameters relaxation, minimization of the free energy (1) with respect to the order parameters, $\psi_C^*$ and $\psi_S^*$, leads to the coupled time-dependent Landau-Ginzburg (TDLG) relaxation equations,

$$-\Gamma \frac{\partial \psi_C}{\partial t} = \frac{\delta F[\psi_C, \psi_S]}{\delta \psi_C^*}, \qquad -\Gamma \frac{\partial \psi_S}{\partial t} = \frac{\delta F[\psi_C, \psi_S]}{\delta \psi_S^*}. \qquad (11)$$

Here the dissipation constant $\Gamma$ is the Landau-Khalatnikov relaxation coefficient [35]. The TDGL equations (11) are used in the finite element modelling (FEM) of the S and C order parameters relaxation from a given initial distribution (e.g., from some regular structure or, more often, randomly small fluctuations) to equilibrium spatially inhomogeneous structures. FEM, as well as more powerful phase-field modeling [36], based on TDLG equations, is a usual tool for numerical search of equilibrium spatially inhomogeneous structures. The approach is widely used for the description of 2D and 3D domain structures of various morphology (stripes, vortices, labyrinths, bubbles, halos, topological defects, etc.) in different materials with coupled long-range orders [36].

Since we are interested in the form of equilibrium structure only, the value of $\Gamma$ does not play any role, when the simulation time $t$ is much higher than the maximal of characteristic relaxation times, $\tau_S = \frac{\Gamma}{|a_s|}$ and $\tau_C = \frac{\Gamma}{|a_c|}$, because initial distributions of the C and S order parameters eventually relax to the equilibrium state. The condition $t \gg max[\tau_S, \tau_C]$ can be fulfilled everywhere except for immediate vicinity of transition temperatures $T_S$ or $T_C$, where $a_s$ or $a_c$ tends to zero. Thus, at low temperatures, considered in this work, the dimensionless "machine" time can be introduced as $\tau = \frac{t}{\tau_{LK}}$, where $\tau_{LK} = max\left[\frac{\Gamma}{|\alpha_S T_S|}, \frac{\Gamma}{|\alpha_C T_C|}\right]$. The explicit form of the coupled equations (11), written in the dimensionless variables, are listed in **Appendix S1** [32].

Using these equations, we analyzed different scenarios of decoupled, coupled and/or intertwined S and C orders; and a schematical illustration some of them is shown in **Fig.3(a), (d)** and **(g).** Results presented in **Fig. 3** correspond to $\tau \gg 10^2$, $\xi_S^* = \xi_C^* = \xi^*$, $v_C^* = v_S^* = v^*$, $w_C^* = w_S^* = w^*$, and $T \to 0$. The top row are profiles calculated at $y = 0$, and the middle and bottom rows are 2D contour maps in $\{x, y\}$ coordinates.

In particular, coexisting homogeneous S order and nonlinear anharmonic CWs is shown in **Fig.3(a)-(c)**. It appeared that the profile of CW is well-described by an elliptic sine function, which phase is apparently decoupled from the magnitude of S order. The anharmonic coupled waves of S and C orders are shown in **Fig.3 (b)-(f)**. It appeared that their profiles, which appeared antiphase, are well-described by a combination of elliptic functions. The localized S-spot, which disturbs the phase of nonlinear CW, is shown in **Fig.3 (g)-(i)**. These examples show that S waves and spots can act as 1D and 2D topological defects for the CW.

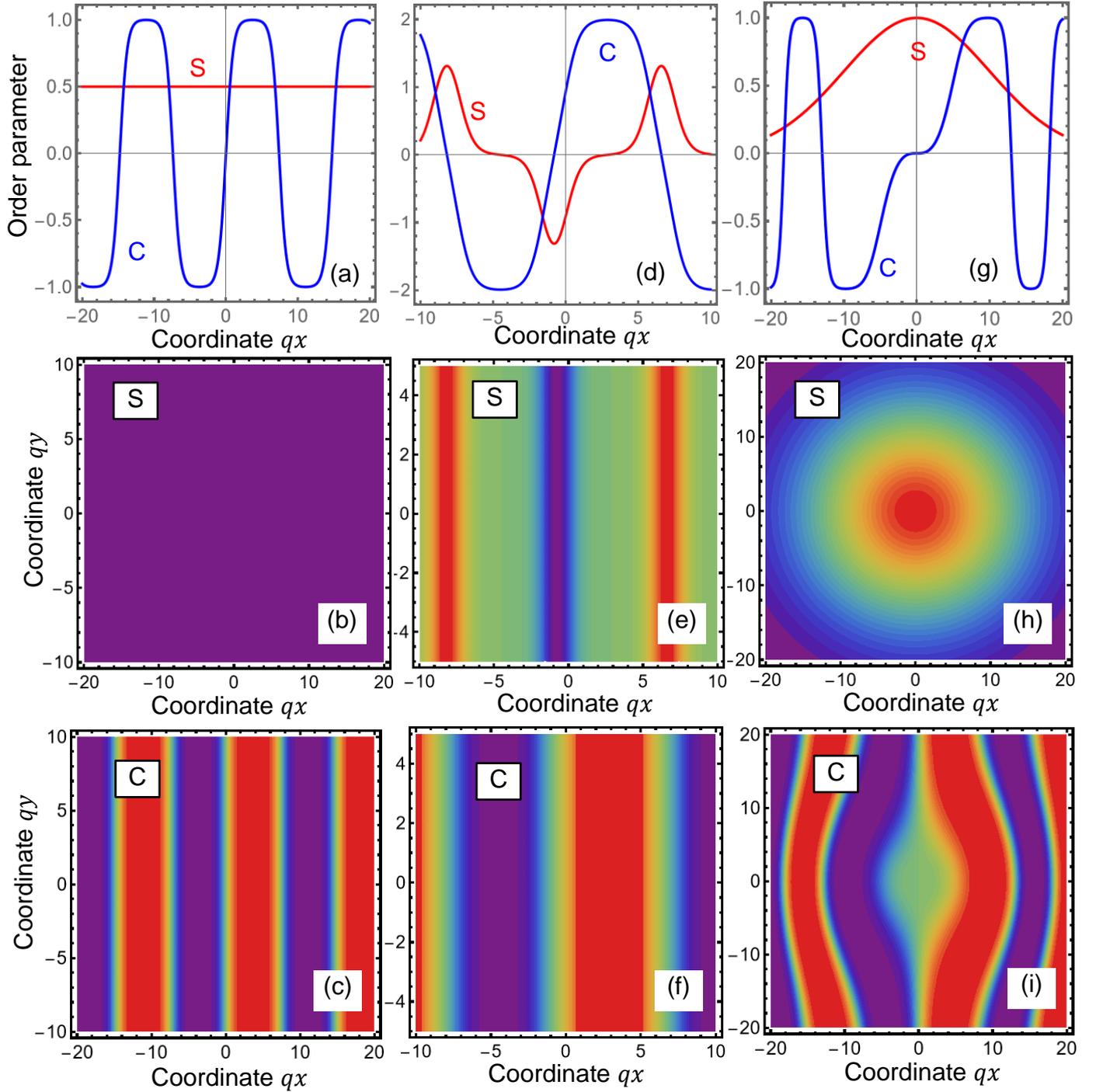

**FIGURE 3.** Schematical illustrations of the coexisting homogeneous S and CW orders **(a)-(c)**; the anharmonic coupled waves of S and C orders **(b)-(f)**; the localized "spot" of S regions, which disturbs the phase of CW **(g)-(i)**. Red curves in the plots (a)-(c) is $\varphi_S$ and blue curves is $\varphi_C$.

## V. CONCLUSIONS

We propose a Landau-Ginzburg description of the charge density waves coupled with lattice and/or electronic long-range ordering in ferroics and/or high-temperature superconductors. We derive analytical expressions for the energies of different phases, corresponding order parameters, waves amplitudes, and modulation periods. Using the analytical expressions, listed in **Table 1**, one can calculate the phase diagrams of versatile ferroics and high-T superconductors with the charge order "C", and the spontaneous long-range (superconductive, polar or magnetic) order parameter "S", corresponding amplitudes and modulation period of charge density waves. The order parameter amplitudes and modulation periods are most sensitive to the biquadratic and biquadratic gradient-coupling strength, which can change their sign, and to the S/C energy ratio, and less sensitive to the positive gradient coefficients. The analytical expressions, obtained in this work, can be employed to guide the comprehensive physical explanation, deconvolution and Bayesian analysis of experimental data on quantum materials ranging from charge-ordered ferroics to high-temperature superconductors.

**Acknowledgements.** We are sincerely grateful to the Referees for stimulating discussions and very useful suggestions. The work is supported by the US Department of Energy, Office of Science, Basic Energy Sciences, under Award Number DE-SC0020145 as part of the Computational Materials Sciences Program. A.N.M. and E.A.E. are also supported by the National Academy of Sciences of Ukraine.

**Authors' contribution.** Research idea belongs to A.N.M. and L.-Q.C. A.N.M. formulated the problem, performed analytical calculations, analyzed results and wrote the manuscript draft. E.A.E. wrote codes and prepared figures. V.G. and L.-Q.C. worked on the manuscript improvement.

**Data availability.** Numerical results presented in the work are obtained and visualized using a specialized software, Mathematica 13.1 [37], and the Mathematica notebook, which contain the codes, is available per reasonable request.

**APPENDIX A. Temperature-dependent phase diagrams**

Results, presented above and in the main text, correspond to very low temperatures, $0 \leq T \ll \min[T_C, T_S]$. The values of $T_C$ and/or $T_S$ can play a very significant role when the temperature $T$ rises, and becomes comparable with $\min[T_C, T_S]$. We hope to consider this question in details in near future. Preliminary calculations for higher temperatures, some results of which are presented in **Figs. A1** and **A2**, show that the ratio $T_C / T_S$ has a principal influence on the form of the phase diagrams. The phase diagrams are very sensitive to the parameters, $\eta^*$ and $\vartheta$ (see **Fig. A1**), relatively sensitive to the parameter $\xi^*$ (see **Fig. A2**, the top row), and almost insensitive to the positive parameter $\chi^*$ (see **Fig. A2**, the bottom row) for given values of $v^*$ and $w^*$.

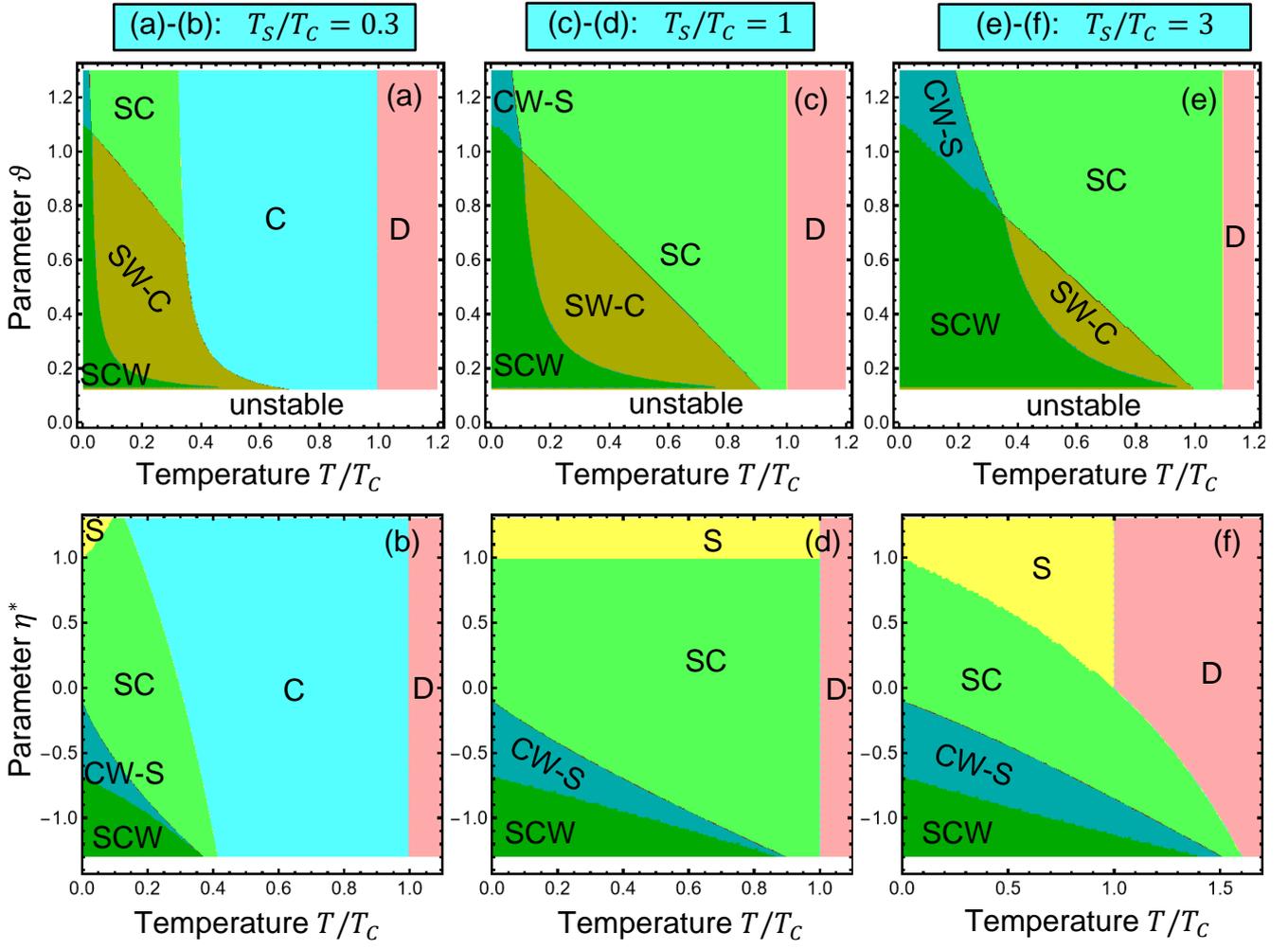

**FIGURE A1.** Phase diagrams in coordinates $T - \vartheta$ calculated for $\eta^* = -0.15$ (**a**, **c**, **e**). Phase diagrams in coordinates $T - \eta^*$ calculated for $\vartheta = 2$ (**b**, **d**, **f**). Other parameters: $T_S/T_C = 0.3$ (**a**, **b**), $T_S/T_C = 1$ (**c**, **d**), $T_S/T_C = 3$ (**e**, **f**); $\xi^* = -1$, $\chi^* = 1$, $\nu^* = 10$, and $w^* = 0.1$.

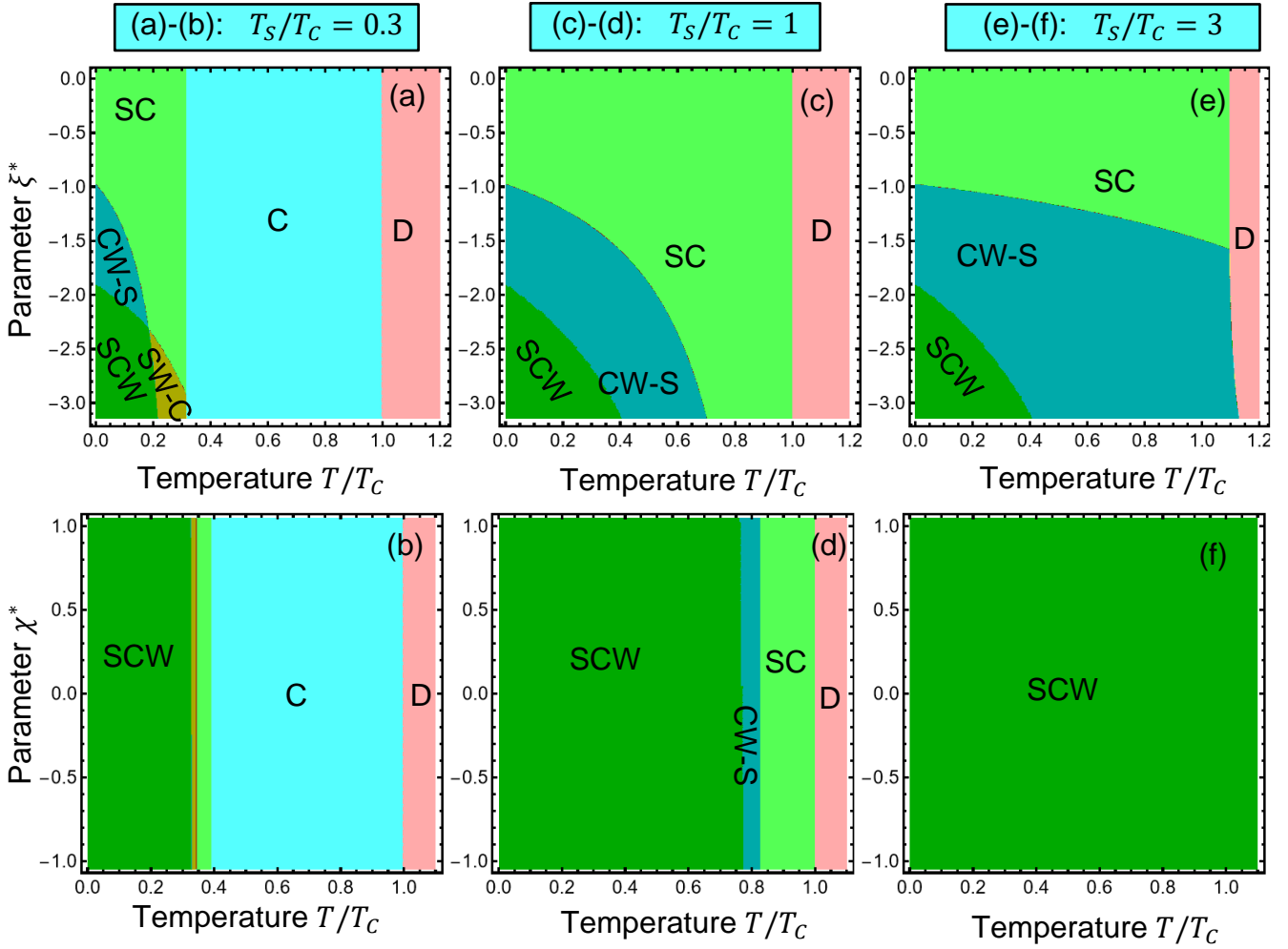

**FIGURE A2.** Phase diagrams in coordinates $T - \xi^*$ calculated for $\chi^* = 1, \eta^* = -0.15$ and $\vartheta = 2$ (**a, c, e**). Phase diagrams in coordinates $T - \chi^*$ calculated for $\vartheta = 2$, and $\eta^* = -1$ $\xi^* = -2$ (**b, d, f**). Other parameters: $T_S/T_C = 0.3$ (**a, b**), $T_S/T_C = 1$ (**c, d**), $T_S/T_C = 3$ (**e, f**); $\nu^* = 10$, and $w^* = 0.1$.

## APPENDIX B. Estimates of biquadratic coupling coefficients

Our model is still far from the quantitative description of available experiments for SCW in high-temperature superconductors, because too many LG parameters is unknown and should be determined from microscopic, e.g., ab initio, calculations. Our model may be suitable for the semi-quantitative description of CW in ferroics, such as an electronic ferroelectric LuFe$_2$O$_4$ [38], transition-metal dichalcogenide TiSe$_2$ [39, 40] with charge density waves and antiferroelectric instability, and rare-earth tritelluride LaTe$_3$ with light-induced "crossed" charge density waves along the a-axis and c-axis [41], as well as for the description of strongly coupled elastic strain, ferroelectric, and/or superconducting orderings in a *R*-doped incipient ferroelectric SrTiO$_3$ [42, 43] (*R* = Sm, La, Nb ...). For ferroelectric and piezoelectric materials, such as LuFe$_2$O$_4$, low-dimensional TiSe$_2$, and strained SrTiO$_3$, the spontaneous polar and/or antipolar order parameters are either measured and/or calculated by DFT [44], or can be estimated from e.g., piezoelectric reaction. For rare-earth tritellurides, such as LaTe$_3$, the transition temperatures, $T_{C1}$ and $T_{C2}$, and wavenumbers $q_{C1}$ and $q_{C2}$ have been measured. Also, the electronic properties of these materials are relatively well-known.

The gradient coefficients $g_C$ and $g_S$ can be estimated from the width of domain walls and/or CW modulation periods. Higher order gradient coefficients, $w_C$ and $w_S$, $v_C$ and $v_S$, can be estimated from modulation period, and it is reasonable for them to be pairwise equal. The poorly known are biquadratic coupling parameters, $\eta$, $\xi$ and $\chi$, which also should be determined (or at least estimated) from available experiments and/or microscopic calculations. To the best of our knowledge, their direct measurements are hardly possible for materials with CW, and we dare to list our indirect estimates in **Table B2**, where the scattering of parameters range are very high and well overlap the ranges of dimensionless parameters shown in **Figures B1-B2** in **Appendix B**.

Here we regard that the long-range parameter $\psi_S$ is a spontaneous (or incipient) polarization $P_S$, measured in C/m². The charge density $\rho$, is normalized as $\rho = \rho_0 \text{Re}[\psi_C]$ [1], where $\rho_0$ is a homogeneous carrier density (in C/m³), and $\psi_C$ is a dimensionless charge-order parameter. The estimates of $\eta$, $\xi$ and $\chi$ are based on the linear relationship [45] between the free change density variation $\delta\rho$, deformation potential [46] and/or Vegard strain [47] tensor $\Xi_{ij}$, and elastic stress variation $\delta u_{ij}$, which role is principally important in ferroelastics like SrTiO$_3$ (see e.g., Ref.[12]) and transition-metal dichalcogenides (see e.g., Refs.[3, 48] and refs. therein). Namely, using the relations $\delta u_{ij} \cong \Xi_{ij}\frac{\delta\rho}{e}$ [45] and $\delta\rho = \rho_0 \text{Re}[\psi_C]$ [1] in the biquadratic coupling contribution, $W_{ijklmn}u_{ij}u_{kl}P_m P_n$ (also known as nonlinear electrostriction coupling [49, 50]), we can estimate that

$$W_{ijklmn}u_{ij}u_{kl}P_m P_n \cong W_{ijklss}\Xi_{ij}^{-1}\Xi_{kl}^{-1} e^2 \delta\rho^2 P_S^2 \cong W_{ijklss}\Xi_{ij}^{-1}\Xi_{kl}^{-1}\frac{e^2}{\rho_0^2}|\psi_C|^2|\psi_S|^2. \quad \text{(B.1a)}$$

Thus

$$\eta \cong W_{ijklss}\Xi_{ij}^{-1}\Xi_{kl}^{-1}\frac{e^2}{\rho_0^2}. \quad \text{(B.1b)}$$

Using the values of parameters $|W_{ijklss}|\sim 10^{11}$ J·m/C² [49, 50], $\Xi \cong (10^{-29}-10^{-30})$ m³ [46, 47], $\frac{\rho_0}{e} \cong 10^{26}$ m⁻³, and $e = 1.6 \cdot 10^{-19}$ C, we obtained that $\eta \cong (10^{17}-10^{19})$·J·m/C². The higher-order gradient coupling coefficients, $\xi$ and $\chi$ are estimated using the characteristic width or uncharged domain walls $d\sim 10^{-9}$ m [12] and wavevector $q_{CW}\sim 10^{+9}$ 1/m [41].

**Table B2.** Phenomenological parameters of the multiparametric LG model

| Phenomenological parameter in the free energy (1) | Corresponding dimensionless parameter(s) in the free energy (7) | Actual range of dimensionless parameter(s) | Magnitude range of a dimension parameters * for a reference materials SrTiO$_3$ and TiSe$_2$ |
|---|---|---|---|
| Coupling constant $\eta$ | $\eta^*$ | $-1.5 - +1.5$ | $\eta = (10^{17}-10^{19})$·J·m/C² |
| gradient coupling $\xi$ | $\xi_s^*$ and $\xi_c^*$, which are assumed to be equal | $-3 - +3$ | $\xi \cong (10^{-1}-10^{+1})$ J·m³/C² |
| gradient coupling $\chi$ | $\chi^*$ | $-1 - +1$ | $\chi \cong (10^{-19}-10^{-17})$ J·m⁵/C² |
| gradient coefficients $g_C$ and $g_S$ | wavevectors | variable | $g_S = (10^{-10}-10^{-11})$·J m³/C² |

| | $q_c = \sqrt{\frac{g_c}{\alpha_c}} q$ and $k_s = \sqrt{\frac{g_s}{\alpha_s}} k$ | | |
|---|---|---|---|
| higher order gradient coefficients $w_C$ and $w_S$ | $w_c^*$ and $w_s^*$, which are assumed to be equal | 0 – 10 | $w_S \cong (10^{-8}$–$10^{-10})$ J·m$^{11}$/C$^2$ |
| higher order gradient coefficients $v_C$ and $v_S$ | $v_c^*$ and $v_s^*$, which are assumed to be equal | 0 – 10 | $v_S \cong (10^{-27}$–$10^{-29})$ J·m$^5$/C$^2$ |
| expansion coefficients $a_C(T) = \alpha_C \left(\frac{T}{T_C} - 1\right)$ and $a_S(T) = \alpha_S \left(\frac{T}{T_S} - 1\right)$ | $\theta_C(T, q_c) = \frac{T}{T_C} - 1 + q_c^2$ and $\theta_S(T, k_s) = \frac{T}{T_S} - 1 + k_s^2$ | functions of temperature and wavevectors; $0.1 \leq \frac{T_S}{T_C} \leq 10$ | $\alpha_S = (10^6$–$10^9)$·J·m/C$^2$ $T_S = (30$–$400)$·K $T_C = (0.4$–$200)$·K |
| higher order expansion coefficients $b_C$ and $b_S$ | $\beta_c(q_c) = 1 + w_c^* q_c^2 + v_c^* q_c^4$ and $\beta_s(k_s) = 1 + w_s^* k_s^2 + v_s^* k_s^4$ | functions of wavevectors | $b_S = (10^9$–$10^{12})$·J·m$^5$/C$^4$ |
| energies $f_c = \frac{\alpha_C^2}{b_C}$ and $f_s = \frac{\alpha_S^2}{b_S}$ | ratio $\vartheta = \frac{\alpha_S^2 b_C}{\alpha_C^2 b_S}$ | 0 – 10 | $f_s \sim (10^5$–$10^7)$·J /(K$^2$·m$^3$) |

\* Unfortunately, we did not find any reliable data for the values of $\alpha_C$ and $b_C$ (and so for $f_C$). Because of this the order parameter $\psi_C$ is selected dimensionless from the beginning, and have an order of unity; its LG parameters, $\alpha_C$ and $b_C$, have the same dimension, J/m$^3$. The gradient coefficients $g_C$, $w_C$ and $v_C$ have a dimension J/m, J/m and J·m, respectively. The characteristic gradient scale is $1/q_{CW} \sim 10^{-9}$ m.

# Supplementary Materials to the paper

**"Landau-Ginzburg theory of charge density wave formation accompanying lattice and electronic long-range ordering"**

by

Anna N. Morozovska[1*], Eugene A. Eliseev[2], Venkatraman Gopalan[3†] and Long-Qing Chen[3‡]

[1] Institute of Physics, National Academy of Sciences of Ukraine,

41, pr. Nauki, 03028 Kyiv, Ukraine

[2] Institute for Problems of Materials Science, National Academy of Sciences of Ukraine,

3, Krjijanovskogo, 03142 Kyiv, Ukraine

[3] Department of Materials Science and Engineering,

Pennsylvania State University, University Park, PA 16802, USA

## Appendix S1. Landau-Ginzburg free energy

### S1.1. Thermodynamically stable phases of the free energy

The considered free energy (1) in a one-component and one-dimensional case is:

$$f[\psi_S, \psi_C] = \alpha_S \left(\frac{T}{T_S} - 1\right)|\psi_S|^2 + b_S|\psi_S|^4 + g_S \left|\frac{\partial \psi_S}{\partial x}\right|^2 + \left(w_S \left|\frac{\partial \psi_S}{\partial x}\right|^2 + v_S \left|\frac{\partial^2 \psi_S}{\partial x^2}\right|^2\right)|\psi_S|^2 +$$

$$\alpha_C \left(\frac{T}{T_C} - 1\right)|\psi_C|^2 + b_C|\psi_C|^4 + g_C \left|\frac{\partial \psi_C}{\partial x}\right|^2 + \left(w_C \left|\frac{\partial \psi_C}{\partial x}\right|^2 + v_C \left|\frac{\partial^2 \psi_C}{\partial x^2}\right|^2\right)|\psi_C|^2 + \eta|\psi_S|^2|\psi_C|^2 +$$

$$\xi\left(|\psi_S|^2 \left|\frac{\partial \psi_C}{\partial x}\right|^2 + |\psi_C|^2 \left|\frac{\partial \psi_S}{\partial x}\right|^2\right) + \chi \left|\frac{\partial \psi_C}{\partial x}\right|^2 \left|\frac{\partial \psi_S}{\partial x}\right|^2. \quad \text{(S.1a)}$$

Expression (S.1a) depends on 15 parameters. At the same time, we cannot decrease the number of parameters due to the following reasons. The 2-4 power LG description of a single temperature-dependent scalar long-range order parameter $\psi_S$, corresponding to the system with a spatially-inhomogeneous phase, requires not less than four parameters, $f_S[\psi_S] = \alpha_S \left(\frac{T}{T_S} - 1\right)|\psi_S|^2 + b_S|\psi_S|^4 + g_S \left|\frac{\partial \psi_S}{\partial x}\right|^2$, where $\alpha_S$, $T_S$, $b_S$ and $g_S$ are the parameters. Since $\psi_S$ interacts with at least one long-range order, namely, with the charge order parameter, $\psi_C$, one need to account for another four fitting parameters, $\alpha_C$, $T_C$, $b_C$ and $g_C$ in $f_C[\psi_C]$. Also at least one coupling coefficient $\eta$, which determines the strength of e.g., biquadratic coupling energy, $\eta|\psi_S|^2|\psi_C|^2$, is required. Thus, the minimal number of independent parameters, which should be included in the LG model of the considered S-C system, is **nine** (see e.g., Eqs.(S.1) and comments to it in the "Supplementary Note 5: Ginzburg-Landau formalism of two competing orders" in Ref.[1]). To stabilize incommensurate spatially-modulated phases of type II, higher-order gradient terms should be mandatory included [2], which requires to add more than **one parameter** in each of the free energy contributions $\psi_S$, $\psi_C$, and $\psi_{int}$. Therefore, the scalar free energy (S.1) and its tensorial form (S.1) include more than **11 parameters**. However, for complete description of the modulated phases the structure of nonlinear higher-order gradient coupling terms are related with corresponding Lifshitz invariants, which leads to addition of at least **2 parameters** in the free energy contributions $\psi_S$, $\psi_C$, and $\psi_{int}$. In result we have $9 + 6 = 15$ parameters.

---

[*] anna.n.morozovska@gmail.com

[†] vgopalan@psu.edu

[‡] lqc3@psu.edu

Within a harmonic approximation and one-dimensional case, the main contributions of the free energy (S.1a) acquire the form, $f_C = (a_C + g_C q^2)|\delta\psi_C|^2 + (b_C + w_C q^2 + v_C q^4)|\delta\psi_C|^4$, $f_S = (a_S + g_S k^2)|\delta\psi_S|^2 + (b_S + w_S k^2 + v_S k^4)|\delta\psi_S|^4$ and $f_{int} = [\eta + \xi(k^2 + q^2) + \chi k^2 q^2]|\delta\psi_S|^2|\delta\psi_C|^2$. So that the free energy (S.1a) acquires the form:

$$f = (a_C + g_C q^2)|\delta\psi_C|^2 + (b_C + w_C q^2 + v_C q^4)|\delta\psi_C|^4 + (a_S + g_S k^2)|\delta\psi_S|^2 + (b_S + w_S k^2 + v_S k^4)|\delta\psi_S|^4 + [\eta + \xi(k^2 + q^2) + \chi k^2 q^2]|\delta\psi_S|^2|\delta\psi_C|^2. \quad (S.1b)$$

Thermodynamically stable homogeneous and spatially-modulated phases of the free energy (S.1b), corresponding absolute values of the order parameters, energies and stability conditions are listed in **Table S1**.

**Table S1.** Thermodynamically stable phases of the free energy (S.1)

| Phase description and abbreviation | Order parameters | Free energy density | Stability conditions (at $b_C > 0$, $b_S > 0$, $2\sqrt{b_C b_S} + \eta > 0$) |
|---|---|---|---|
| Disordered (**D**) | $\psi_C = \psi_S = 0$ | 0 | $a_C > 0$, $a_S > 0$ |
| Spatially-homogeneous scalar order (**S**) | $|\psi_C| = 0$, $|\psi_S| = \sqrt{-\frac{a_S}{2b_S}}$ | $f_S = -\frac{a_S^2}{4b_S}$ | $f_S = min$, $a_S < 0$, $2a_C b_S - \eta a_S > 0$ |
| Spatially-homogeneous mixed (**SC**) | $|\psi_C| = \sqrt{-\frac{2a_C b_S - \eta a_S}{4b_C b_S - \eta^2}}$, $|\psi_S| = \sqrt{-\frac{2a_S b_C - \eta a_C}{4b_C b_S - \eta^2}}$ | $f_{SC} = \frac{-b_S a_C^2 - b_C a_S^2 + \eta a_S a_C}{4b_C b_S - \eta^2}$ | $f_{SC} = min$, $2a_C b_S - \eta a_S < 0$, $2a_S b_C - \eta a_C < 0$, $4b_C b_S > \eta^2$ |
| Spatially-homogeneous charge order (**C**) | $|\psi_C| = \sqrt{-\frac{a_C}{2b_C}}$, $\psi_S = 0$ | $f_C = -\frac{a_C^2}{4b_C}$ | $f_C = min$, $a_C < 0$, $2a_S b_C - \eta a_C > 0$ |
| Spatially-modulated phase with coexisting S and C orders (**SCW**) | In harmonic approximation $\psi_C = \delta\psi_C \exp(\pm iqx)$, $\psi_S = \delta\psi_S \exp(\pm ikx)$ The amplitudes $\delta\psi_C$ and $\delta\psi_S$ are given by Eqs.(S.2) | Expression for $f_{SCW}$ is given by Eq.(S.3). Analytical expressions for $k$ and $q$ exist in special cases | $f_{SCW} = min$, $2a_C b_S + \eta a_S < 0$, $2a_S b_C + \eta a_C < 0$, $\xi < 0$ |

Note, that if one considers a single (real or complex) order parameter, $\psi_C$, and include its quadratic and the fourth power terms in the Landau free energy, $f_C = a_C |\psi_C|^2 + b_C |\psi_C|^4$, such energy can describe the second order phase transitions only. When one adds another order parameter, $\psi_S$, and corresponding $f_S =$

$a_S|\psi_S|^2 + b_S|\psi_S|^4$, and a biquadratic coupling energy, $f_{int} = \eta|\psi_S|^2|\psi_C|^2$, the total free energy, $f = f_S + f_C + f_{int}$, may contain a line of the first order phase transitions. This can be readily checked by the analysis of all solutions of the coupled equations for the order parameters, which are listed in **Table S1**.

Under the condition $-2\sqrt{b_C b_S} < \eta < 2\sqrt{b_C b_S}$ three stable ordered phases, C, S and SC, are possible. The transition between the S phase and the SC phase is determined by the condition $2a_C b_S = \eta a_S$. The transition between the C phase and the SC phase is determined by the condition $2a_S b_C = \eta a_C$. Under the conditions, $-2\sqrt{b_C b_S} < \eta < 2\sqrt{b_C b_S}$, these transitions are continuous (i.e., the transition between S- (or C-) and SC-phases is of the second order).

In contrast, under the condition $2\sqrt{b_C b_S} < \eta$, the SC phase is unstable and the transition between the stable S phase and C phase is determined by the energy balance, $f_S = f_C$, meanwhile the conditions $2a_C b_S = \eta a_S$ and $2a_S b_C = \eta a_C$ determine the boundaries of the coexistence region of the S and C phases. Corresponding transition point, $f_S = f_C$, is within the region and has a discontinuous nature, because $|\psi_S|$ has a jump $\sqrt{-\frac{a_S}{2b_S}}$ for $a_S < 0$, and $|\psi_C|$ has a jump $\sqrt{-\frac{a_C}{2b_C}}$ for $a_C < 0$ (i.e., the transition between S and C phases is of the first order).

The minimization of Eq.(S.1) in the SCW phase yields:

$$|\delta\psi_C| = \sqrt{-\frac{2(a_C+g_C q^2)(b_S+w_S k^2+v_S k^4)+[\eta+\xi(k^2+q^2)+\chi k^2 q^2](a_S+g_S k^2)}{4(b_C+w_C q^2+v_C q^4)(b_S+w_S k^2+v_S k^4)-[\eta+\xi(k^2+q^2)+\chi k^2 q^2]^2}}, \quad (S.2a)$$

$$|\delta\psi_S| = \sqrt{-\frac{2(a_S+g_S k^2)(b_C+w_C q^2+v_C q^4)+[\eta+\xi(k^2+q^2)+\chi k^2 q^2](a_C+g_C q^2)}{4(b_C+w_C q^2+v_C q^4)(b_S+w_S k^2+v_S k^4)-[\eta+\xi(k^2+q^2)+\chi k^2 q^2]^2}}. \quad (S.2b)$$

The energy minimum is

$$f_{SCW} = -\frac{(b_S+w_S k^2+v_S k^4)(a_C+g_C q^2)^2+(b_C+w_C q^2+v_C q^4)(a_S+g_S k^2)^2}{4(b_C+w_C q^2+v_C q^4)(b_S+w_S k^2+v_S k^4)-[\eta+\xi(k^2+q^2)+\chi k^2 q^2]^2} +$$

$$\frac{[\eta+\xi(k^2+q^2)+\chi k^2 q^2](a_S+g_S k^2)(a_C+g_C q^2)}{4(b_C+w_C q^2+v_C q^4)(b_S+w_S k^2+v_S k^4)-[\eta+\xi(k^2+q^2)+\chi k^2 q^2]^2}. \quad (S.3)$$

Since $b_{SC} > 0$ and $b_{SC} > 0$, the energy $f_{SCW}$ can diverge if:

$$4(b_C + w_C q^2 + v_C q^4)(b_S + w_S k^2 + v_S k^4) - [\eta + \xi(k^2 + q^2) + \chi k^2 q^2]^2 = 0. \quad (S.4)$$

Under the conditions $4b_C b_S - \eta^2 > 0, w_C > 0, w_S > 0, \chi > 0$ and $k^2 + q^2 \geq 0$, the equality (S.4) is possible if $\xi > 0$. So, the appearance of SCW instability is possible at $\xi < 0$.

### S1.2. The dimensionless variables and order parameters in the free energy (S.1)

Let us introduce the dimensionless variables and order parameters in the free energy (S.1b):

$$\varphi_C^2 = \frac{|\psi_C|^2}{\psi_{C0}^2}, \quad \varphi_S^2 = \frac{|\psi_S|^2}{\psi_{S0}^2}, \quad \psi_{C0}^2 = \frac{\alpha_C}{2b_C}, \quad \psi_{S0}^2 = \frac{\alpha_S}{2b_S}, \quad f_C = \frac{\alpha_C^2}{b_C}, \quad f_S = \frac{\alpha_S^2}{b_S}, \quad \vartheta = \frac{f_S}{f_C}. \quad (S.5a)$$

Dimensionless wavenumbers and positive gradient coefficients

$$q_C = \sqrt{\frac{g_C}{\alpha_C}} q, \quad\quad k_S = \sqrt{\frac{g_S}{\alpha_S}} k, \quad (S.5b)$$

$$v_C^* = \frac{v_C}{b_C}\left(\frac{\alpha_C}{g_C}\right)^2, \quad v_S^* = \frac{v_S}{b_S}\left(\frac{\alpha_S}{g_S}\right)^2, \quad w_C^* = \frac{w_C}{b_C}\frac{\alpha_C}{g_C}, \quad w_S^* = \frac{w_S}{b_S}\frac{\alpha_S}{g_S}, \quad \text{(S.5c)}$$

which contribute to the temperature- and wavenumber- dependent parameters:

$$\theta_C(T, q_c) = \frac{T}{T_C} - 1 + q_c^2, \qquad \theta_S(T, k_s) = \frac{T}{T_S} - 1 + k_s^2, \quad \text{(S.6a)}$$

$$\beta_C(q_c) = 1 + w_C^* q_c^2 + v_C^* q_c^4, \qquad \beta_S(k_s) = 1 + w_S^* k_s^2 + v_S^* k_s^4, \quad \text{(S.6b)}$$

and coupling constant

$$\mu(k_s, q_c) = \frac{\alpha_C \alpha_S}{2b_C b_S f_C}[\eta + \xi(k^2 + q_c^2) + \chi k^2 q_c^2] \equiv \eta^* + \xi_s^* k_s^2 + \xi_c^* q_c^2 + \chi^* k_s^2 q_c^2, \quad \text{(S.6c)}$$

$$\eta^* = \frac{\alpha_C \alpha_S}{2 b_C b_S}\frac{\eta}{f_C}, \quad \xi_s^* = \frac{\alpha_C \alpha_S}{2 b_C b_S}\frac{\xi}{f_C}\frac{\alpha_S}{g_S}, \quad \xi_c^* = \frac{\alpha_C \alpha_S}{2 b_C b_S}\frac{\xi}{f_C}\frac{\alpha_C}{g_C}, \quad \chi^* = \chi\frac{(\alpha_C \alpha_S)^2}{2 b_C b_S}\frac{\xi}{f_C g_C g_S}. \quad \text{(S.6d)}$$

Using the dimensionless variables (S.6) and order parameters (S.5), we obtain from Eq.(S.1) that:

$$\frac{f}{f_C} = \left[\theta_C(T, q_c)\frac{\varphi_C^2}{2} + \beta_C(q_c)\frac{\varphi_C^4}{4}\right] + \vartheta\left[\theta_S(T, k_s)\frac{\varphi_S^2}{2} + \beta_S(k_s)\frac{\varphi_S^4}{4}\right] + \mu(k_s, q_c)\frac{\varphi_C^2 \varphi_S^2}{2}. \quad \text{(S.7)}$$

The free energy (S.7) is stable at high values of the order parameters and wavenumbers under the simultaneous validity of the conditions $\eta^* > -\sqrt{\vartheta}$, $w_S^* > -\sqrt{4 v_S^*}$, $w_C^* > -\sqrt{4 v_C^*}$, $\xi_c^* > -\sqrt{v_C^*}$, $\xi_s^* > -\sqrt{\vartheta v_S^*}$, $\chi^* > -\sqrt{\vartheta v_C^* v_S^*}$, as well as $2 v_S^* \eta^* > w_S^* \xi_s^* - \sqrt{(4 v_S^* - w_S^{*2})(\vartheta v_S^* - \xi_s^{*2})}$, and $2 v_C^* \eta^* > w_C^* \xi_c^* - \sqrt{(4 v_C^* \vartheta - w_C^{*2})(v_C^* - \xi_c^{*2})}$. Note that the last two conditions become redundant if $w_S^* > 0$ and $w_C^* > 0$. Note that the parameter $\vartheta$ can be arbitrary: small, close to unity or big; and these cases are analyzed below.

Thermodynamically stable homogeneous phases and spatially-modulated states of the energy (S.7), corresponding absolute values of the order parameters, phase energies and stability conditions are listed in **Table 1** in the main text. The cumbersome expression for the free energies of $f_{CW-S}$, $f_{SW-C}$ and $f_{SCW}$ are:

$$f_{CW-S} = \frac{-\vartheta^2 w_C^{*2}\theta_S^2 + 4\left[-\eta^{*2} + \vartheta + \vartheta v_C^*(\theta_C^2 - 2\eta^*\theta_C\theta_S + \vartheta\theta_S^2) - \xi_s^*(-2\eta^*\theta_C + 2\vartheta\theta_S + \theta_C^2\xi_s^*)\right] + 4\vartheta w_C^*[\eta^*\theta_S + \theta_C(-1 + \theta_S\xi_s^*)]}{4\left[4(\eta^{*2} - \vartheta)v_C^* + \vartheta w_C^{*2} - 4\eta^* w_C^*\xi_c^* + 4\xi_c^{*2}\right]}, \quad \text{(S.8)}$$

$$f_{SW-C} = \vartheta \frac{-w_S^{*2}\theta_C^2 + 4w_S^*[\theta_C(\eta^* + \theta_S\xi_c^*) - \vartheta\theta_S] + 4\left[\vartheta - \eta^{*2} + v_S^*(\theta_C^2 - 2\eta^*\theta_C\theta_S + \vartheta\theta_S^2) - \xi_c^*(2\theta_C - 2\eta^*\theta_S + \theta_S^2\xi_c^*)\right]}{4\left[4(\eta^{*2} - \vartheta)v_S^* + \vartheta w_S^{*2} - 4\eta^* w_S^*\xi_c^* + 4\xi_c^{*2}\right]}, \quad \text{(S.9)}$$

$$\Delta = 16(\eta^{*2} - \vartheta)\chi^{*2} + \vartheta w_C^*\left(\vartheta w_C^*(-4v_S^* + w_S^{*2}) - 8\chi^*(\eta^* w_S^* - 2\xi_c^*)\right) - 4\vartheta v_C^*\left(4(\eta^{*2} - \vartheta)v_S^* + \vartheta w_S^{*2} - 4\eta^* w_S^*\xi_c^* + 4\xi_c^{*2}\right) + 8(2\eta^*\vartheta v_S^* w_C^* + 2\vartheta\chi^* w_S^* - (4\eta^*\chi^* + \vartheta w_C^* w_S^*)\xi_c^*)\xi_s^* + 16(-\vartheta v_S^* + \xi_c^{*2})\xi_s^{*2}, \quad \text{(S.10a)}$$

$$\Delta_C = 2\vartheta\Bigg(2\vartheta v_C^* w_S^{*2}\theta_C + 4w_S^*(\eta^*\chi^* - \vartheta v_C^*(\eta^* + \theta_S\xi_c^*) + (-2\chi^*\theta_C + \xi_c^*)\xi_s^*) + \vartheta w_C^*\Big(-4\chi^* - w_S^{*2} + 2w_S^*(\chi^*\theta_S + \xi_s^*) + v_S^*(4 - 4\theta_S\xi_s^*)\Big) + 8\Big((-\chi^* + \vartheta v_C^*)\xi_c^* + \eta^*(\chi^* - v_S^*)\xi_s^* - \xi_c^*\xi_s^{*2} + \theta_S(-\eta^*\chi^{*2} + \eta^*\vartheta v_C^* v_S^* + \chi^*\xi_c^*\xi_s^*) + \theta_C(\chi^{*2} + v_S^*(-\vartheta v_C^* + \xi_s^{*2}))\Big)\Bigg), \quad \text{(S.10b)}$$

$$\Delta_q = -\left\{2\vartheta\left(8(\eta^{*2}-\vartheta)\chi^* - 2w_S^*(\vartheta w_S^* + 2\eta^*\chi^*\theta_C - 2\vartheta\chi^*\theta_S) + 8(\eta^* w_S^* + \chi^*\theta_C - \eta^*\chi^*\theta_S)\xi_C^* - 8\xi_C^{*2} + \right.\right.$$
$$\vartheta w_C^*\left(4\xi_C^* + w_S^*(w_S^*\theta_C - 2(\eta^* + \theta_S\xi_C^*))\right)\right) + 4(\vartheta w_S^* - (2\eta^* + w_S^*\theta_C)\xi_C^* + 2\theta_S\xi_C^{*2})\xi_S^* + v_S^*\left(-4\vartheta w_C^*(\theta_C - \right.$$
$$\left.\left.\eta^*\theta_S) + 8(-\eta^{*2} + \vartheta + (\eta^*\theta_C - \vartheta\theta_S)\xi_S^*)\right)\right)\right\}, \quad \text{(S.10c)}$$

$$\Delta_S = -2\vartheta^2 w_C^{*2}(w_S^* - 2v_S^*\theta_S) + 8\vartheta w_S^*(-\chi^* + v_C^*(\vartheta - \theta_C\xi_C^*)) + 4\vartheta w_C^*(2\eta^*\chi^* + w_S^*(\chi^*\theta_C + \xi_C^*) + $$
$$2\xi_C^*(-2\chi^*\theta_S + \xi_S^*) - 2v_S^*(\eta^* + \theta_C\xi_S^*)) + 16\left(\eta^*(\chi^* - \vartheta v_C^*)\xi_C^* + \vartheta\theta_S\left(\chi^{*2} + v_C^*(-\vartheta v_S^* + \xi_C^{*2})\right) - \right.$$
$$\left.(\vartheta\chi^* - \vartheta v_S^* + \xi_C^{*2})\xi_S^* + \theta_C(\eta^*\vartheta v_C^* v_S^* + \chi^*(-\eta^*\chi^* + \xi_C^*\xi_S^*))\right), \quad \text{(S.10d)}$$

$$\Delta_k = 2\left(8(-\eta^{*2} + \vartheta)\chi^* + \vartheta w_C^*(2\eta^*(w_S^* + 2\chi^*\theta_S) + \vartheta w_C^*(2 - w_S^*\theta_S) - 4(\chi^*\theta_C + \xi_C^*)) + \right.$$
$$4\vartheta v_C^*\left(w_S^*(-\eta^*\theta_C + \vartheta\theta_S) + 2(\eta^{*2} - \vartheta + (\theta_C - \eta^*\theta_S)\xi_C^*)\right) + 2(-2\vartheta(w_S^* + 2\chi^*\theta_S) + 4\eta^*(\chi^*\theta_C + \xi_C^*) + $$
$$\left.\vartheta w_C^*(-4\eta^* + w_S^*\theta_C + 2\theta_S\xi_C^*))\xi_S^* + 8(\vartheta - \theta_C\xi_C^*)\xi_S^{*2}\right). \quad \text{(S.10e)}$$

$$f_{SCW} = \left[\theta_C \frac{\varphi_C^2}{2} + (1 + w_c^* q_c^2 + v_c^* q_c^4)\frac{\varphi_C^4}{4}\right] + \vartheta\left[\theta_S \frac{\varphi_S^2}{2} + (1 + w_s^* k_s^2 + v_s^* k_s^4)\frac{\varphi_S^4}{4}\right] + (\eta^* + \xi_s^* k_s^2 + \xi_c^* q_c^2 + $$
$$\chi^* k_s^2 q_c^2)\frac{\varphi_C^2 \varphi_S^2}{2}. \quad \text{(S.10f)}$$

### S1.3. Evident form of the time-dependent Landau-Ginzburg equations

Allowing for the Khalatnikov mechanism of the order parameters relaxation, minimization of the free energy (1) with respect to the order parameters, $\boldsymbol{\psi}_C^*$ and $\boldsymbol{\psi}_S^*$, leads to the coupled time-dependent Landau-Ginzburg (TDLG) relaxation equations,

$$-\Gamma \frac{\partial \psi_C}{\partial t} = \frac{\delta F[\psi_C, \psi_S]}{\delta \psi_C^*}, \qquad -\Gamma \frac{\partial \psi_S}{\partial t} = \frac{\delta F[\psi_C, \psi_S]}{\delta \psi_S^*}. \quad \text{(S.11)}$$

Here $\Gamma$ is the Landau-Khalatnikov relaxation (dissipation) coefficient. Equations (S.11) are used in the finite element modelling (FEM) of the S and C order parameters relaxation from a given initial distribution (e.g., from some regular structure or, more often, randomly small fluctuations) to equilibrium spatially inhomogeneous structures. FEM, as well as more powerful phase-field modeling [3], is a usual tool for numerical search of equilibrium spatially inhomogeneous structures, such as 2D and 3D vortices, stripes, labyrinths, bubbles, halos, and topological defects, in materials with coupled long-range orders.

If we are interested in the form of equilibrium structure only, the value of $\Gamma$ does not play any role, when the simulation time $t$ is much higher than the maximal of characteristic relaxation times, $\tau_S = \frac{\Gamma}{|a_s|}$ and $\tau_C = \frac{\Gamma}{|a_c|}$, because initial distributions of the C and S order parameters eventually relax to the equilibrium state. The condition $t \gg max[\tau_S, \tau_C]$ can be fulfilled everywhere except for immediate vicinity of transition temperatures $T_S$ or $T_C$, where $a_s$ or $a_c$ tends to zero. Thus, at low temperatures, considered in this work, the dimensionless "machine" time can be introduced as $\tau = \frac{t}{\tau_{LK}}$, where $\tau_{LK} = max[\tau_{S0}, \tau_{C0}]$, $\tau_{S0} = \frac{\Gamma}{|\alpha_S T_S|}$ and $\tau_{C0} = \frac{\Gamma}{|\alpha_C T_C|}$.

Assuming that $\tau_{S0} \leq \tau_{C0}$ for the sake of clarity, the TDLG equations (S.11), written in dimensionless variables (S.5), acquire the form:

$$-\frac{\partial \varphi_C}{\partial \tau} = \left(\frac{T}{T_C} - 1\right)\varphi_C + |\varphi_C|^2 \varphi_C - \Delta \varphi_C + w_c^*[\varphi_C |\nabla \varphi_C|^2 - div(|\varphi_C|^2 \nabla \varphi_C)] + v_c^*[\varphi_C |\Delta \varphi_C|^2 +$$
$$\Delta(|\varphi_C|^2 \Delta \varphi_C)] + \eta^* \varphi_C |\varphi_S|^2 + \xi_s^* \varphi_C |\nabla \varphi_S|^2 - \xi_c^* div(|\varphi_S|^2 \nabla \varphi_C) + \chi^* div[\nabla \varphi_C |\Delta \varphi_S|^2], \quad (S.12a)$$

$$-\frac{\tau_{S0}}{\tau_{C0}}\frac{\partial \varphi_S}{\partial \tau} = \left(\frac{T}{T_S} - 1\right)\varphi_S + |\varphi_S|^2 \varphi_S - \Delta \varphi_S + w_s^*[\varphi_S |\nabla \varphi_S|^2 - div(|\varphi_S|^2 \nabla \varphi_S)] + v_s^*[\varphi_S |\Delta \varphi_S|^2 +$$
$$\Delta(|\varphi_S|^2 \Delta \varphi_S)] + \frac{\eta^*}{\vartheta}\varphi_S |\varphi_C|^2 + \frac{\xi_c^*}{\vartheta}\varphi_S |\nabla \varphi_C|^2 - \frac{\xi_s^*}{\vartheta} div(|\varphi_C|^2 \nabla \varphi_S) + \frac{\chi^*}{\vartheta} div[\nabla \varphi_S |\Delta \varphi_C|^2]. \quad (S.12b)$$

Here we regard $g_S = g_C = g$, and introduce the dimensionless coordinates, $\tilde{x} = x/\sqrt{g}$, and $\tilde{y} = y/\sqrt{g}$; $\nabla$ and $\Delta$ are the gradient and Laplace operators, written in the dimensionless coordinates, respectively.

### S1.4. Some analytical results

Since we are mostly interested in the conditions of the spatially-modulated phases stability at corresponding equilibrium wavenumbers, we can take all possible measures to reduce the number of dimensionless parameters in the energy (S.1). After a proper introduction of dimensionless variables, given by expressions (S.5) and (S.6), the parameters $g_C$ and $g_S$ are involved to the variable wavevectors, $q_c = \sqrt{\frac{g_C}{\alpha_C}} q$ and $k_s = \sqrt{\frac{g_S}{\alpha_S}} k$, and the parameters $b_C$ and $b_S$ are involved to the normalization of the spontaneous order parameters. Hence, the dimensionless free energy (S.7) depends on eleven dimensionless parameters. They are two temperature ratios, $\frac{T}{T_C}$ and $\frac{T}{T_S}$, a constant of the S-C biquadratic coupling strength, $\eta^*$, three constants of the biquadratic gradient-coupling strength, $\xi_s^*$, $\xi_c^*$, and $\chi^*$, four gradient coefficients of S and C subsystems, $v_c^*$, $v_s^*$, $w_c^*$, and $w_s^*$, their energy ratio, $\vartheta$. For further reduction of the number of independent parameters up to eight, we analyzed the realistic case $\xi_s^* = \xi_c^* = \xi^*$, $v_c^* = v_s^* = v^*$ and $w_c^* = w_s^* = w^*$.

Putting $T \to 0$, we can exclude two transition temperatures, $T_S$ and $T_C$ from our consideration. As a result, the considered six-parametric LG model is not much more complex than earlier LG models [1] for two coupled long-range orders with respect to the number of fitting parameters.

Relatively simple analytical results can be derived from Eq.(S.7) for a special case

$$\vartheta = 1, v_c^* = v_s^* = v^*, w_c^* = w_s^* = w^*, \xi_s^* = \xi_c^* = \xi^*. \quad (S.13)$$

For the case one can assume that $q_c = k_s = q_*$, $\varphi_C = \varphi_S = \varphi_*$ and the free energy (S.7) acquires a much simpler form:

$$\frac{f}{f_c} = \theta(T, q_*)\varphi_*^2 + [\beta(T, q_*) + \mu(q_*)]\frac{\varphi_*^4}{2}, \quad (S.14a)$$

where

$$\theta(T, q_*) = \frac{T}{T_C} - 1 + q_*^2, \quad \beta(T, q_*) = 1 + w^* q_*^2 + v^* q_*^4, \quad \mu(q_*) = \eta^* + 2\xi^* q_*^2 + \chi^* q_*^4. \qquad (S.14b)$$

The stability conditions are $v^* + \chi^* > 0$ and $\eta^* + 1 \geq 0$. Minimization of Eqs.(S.14a) leads to the pair of equations

$$\frac{\partial f}{\partial q_*} = 2q_* \left[ \frac{\varphi_*^2}{2} + (w^* + 2v^* q_*^2 + 2\xi^* + 2\chi^* q_*^2) \frac{\varphi_*^4}{4} \right] = 0, \qquad (S.15a)$$

$$\frac{\partial f}{\partial \varphi_*} = \left[ \frac{T}{T_C} - 1 + q_*^2 + (1 + w^* q_*^2 + v^* q_*^4 + \eta^* + 2\xi^* q_*^2 + \chi^* q_*^4) \varphi_*^2 \right] \varphi_* = 0. \qquad (S.15b)$$

Nonzero solution of Eqs.(S.15) is:

$$\varphi_*^2 = \frac{1 - \frac{T}{T_C} - q_*^2}{1 + \eta^* + (w^* + 2\xi^*) q_*^2 + (v^* + \chi^*) q_*^4}, \qquad (S.16a)$$

$$-\frac{\frac{T}{T_C} - 1 + q_*^2}{1 + \eta^* + (w^* + 2\xi^*) q_*^2 + (v^* + \chi^*) q_*^4} = -\frac{2}{w^* + 2\xi^* + (2v^* + 2\chi^*) q_*^2}. \qquad (S.16b)$$

The Eqs.(S.16b) is a biquadratic equation:

$$1 + \eta^* + \frac{w^* + 2\xi^*}{2} \left( \frac{T}{T_C} - 1 \right) + \left( 3 \frac{w^* + 2\xi^*}{2} + (v^* + \chi^*) \left( \frac{T}{T_C} - 1 \right) \right) q_*^2 + 2(v^* + \chi^*) q_*^4 = 0, \qquad (S.17a)$$

which solution is

$$q_*^2 = \frac{ -\left[ 3 \frac{w^* + 2\xi^*}{2} + (v^* + \chi^*) \left( \frac{T}{T_C} - 1 \right) \right] \pm \sqrt{ \left[ 3 \frac{w^* + 2\xi^*}{2} + (v^* + \chi^*) \left( \frac{T}{T_C} - 1 \right) \right]^2 - 8(v^* + \chi^*) \left[ 1 + \eta^* + \frac{w^* + 2\xi^*}{2} \left( \frac{T}{T_C} - 1 \right) \right] } }{ 4(v^* + \chi^*) } \qquad (S.17b)$$

The SCW state is stable when its energy, $f_{SCW}(k_s, q_c)$, is minimal is comparison with the energies of homogeneous phases, $f_C = -\frac{1}{4} \left( \frac{T}{T_C} - 1 \right)^2$ and $f_S = -\frac{1}{4} \left( \frac{T}{T_S} - 1 \right)^2$.

## Appendix S2. Free energy relief

The potential landscape of the free energy (S.7) is analyzed in **Figs. S1-S3.** Using the LG approach [4], where a similar, but simpler free energy was analyzed, the free energy (S.3) as a function of order parameter amplitudes, $|\varphi_C|$ and $|\varphi_S|$, is shown in **Fig. S1** for $\vartheta = 1$, different $\mu$ values, and low temperatures allowing us to assume that $\theta_C \approx \theta_S \approx -1$. Two symmetry-different homogeneous phases exist, namely:

1) The mixed SC phase with the minimal energy density $f = -\frac{1}{2(1+\mu)}$ corresponds to the order parameters $|\varphi_C| = |\varphi_S| = \frac{1}{\sqrt{1+\mu}}$ (see **Figs. S1a-c**). The SC phase is stable at $-1 < \mu < 1$.

2) Coexisting S and C phases with the minimal and energy density $f = -\frac{1}{4}$ corresponds to the order parameters $|\varphi_S| = 1$, $|\varphi_C| = 0$, or $|\varphi_S| = 0$, $|\varphi_C| = 1$ (see **Figs. S1e-f**). S and C phases are stable at $\mu > 1$. The transition to coexisting S and C phases takes place at $\mu = 1$, when all four potential minima merge and transform into a circle (see **Figs. S1d**).

The free energy (S.7) as a function of order parameter amplitudes, $\varphi_C$ and $\varphi_S$, is shown in **Fig. S2** for high value $\vartheta = 10$, different $\mu$ values, and $\theta_C \approx \theta_S \approx -1$. The mixed SC phase is stable for negative and small positive $\mu$ values, namely for $\mu < 0.7$. The single S phase is stable for higher positive $\mu$ values, namely for $\mu > 0.7$.

The free energy (S.7) as a function of order parameter amplitudes, $\varphi_C$ and $\varphi_S$, is shown in **Fig. S3** for small value $\vartheta = 0.1$, different $\mu$ values, and $\theta_C \approx \theta_S \approx -1$. The mixed SC phase is stable for negative and very small positive $\mu$ values, namely for $\mu < 0.1$. The single C phase is stable for $\mu \geq 0$.

The common feature of the free energy (S.7) dependence on the order parameter amplitudes, shown in **Figs. S1-S3**, is the appearance of the stable SC phase at $\mu \leq 0$ independently on $\vartheta$ values. The coexistence of the S and C phases with relatively deep energy minima is possible only for $\vartheta = 1$ and $\mu > 1$; the stable S or C phase with deep energy minima appears at $\mu > 0$, and $\vartheta \gg 1$ or $\vartheta \ll 1$, respectively.

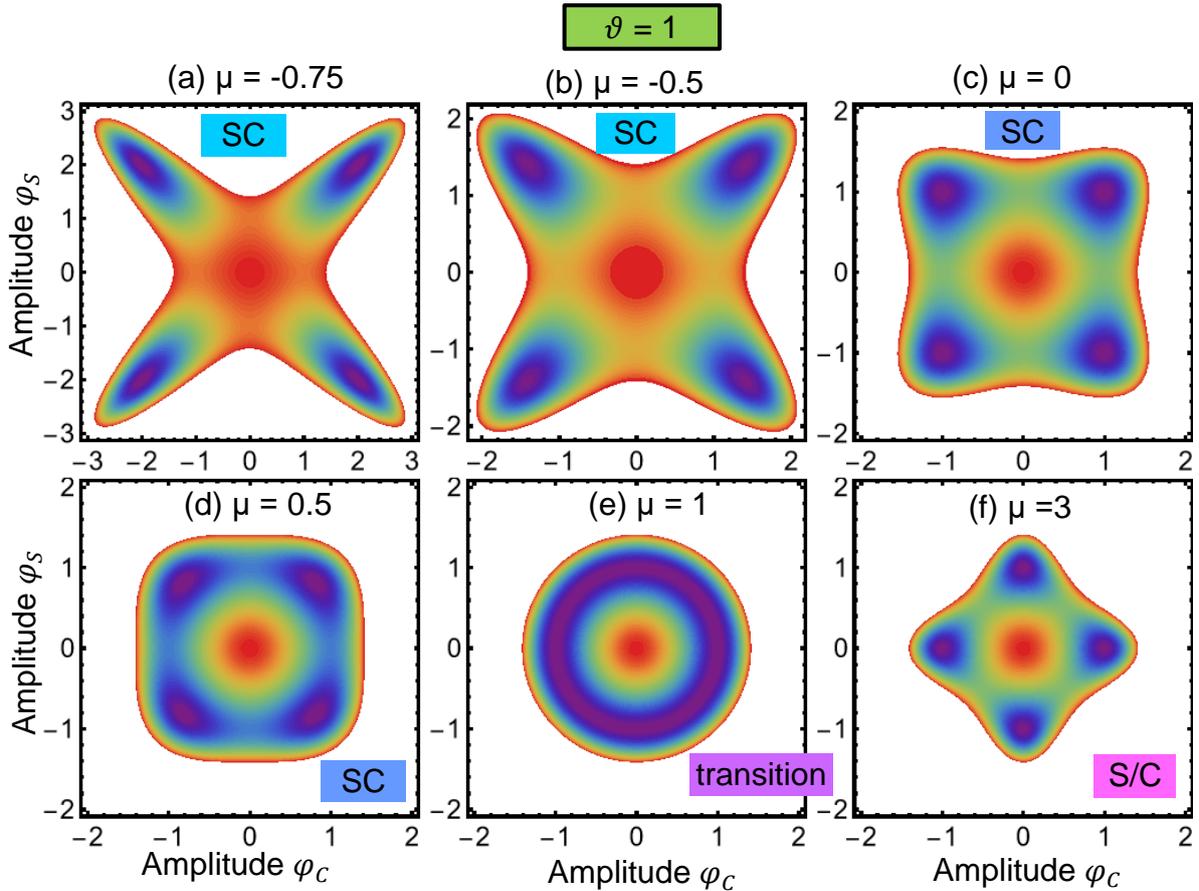

**FIGURE S1.** The free energy (S.7), as a function of order parameter amplitudes $\varphi_C$ and $\varphi_S$, calculated for different values of parameter $\mu$: **(a)** $\mu = -0.75$, **(b)** $\mu = -0.5$, **(c)** $\mu = 0$, **(d)** $\mu = 0.5$, **(e)** $\mu = 1$ and **(f)** $\mu = 3$. Red color denotes zero energy, while violet color is its minimal density in relative units. Capital letters "SC", and "S/C" denote the mixed spontaneous long-range order - charge order state, coexisting spontaneous long-range order and charge order, respectively. Other parameters: $\vartheta = 1$ and $\theta_C \approx \theta_S \approx -1$.

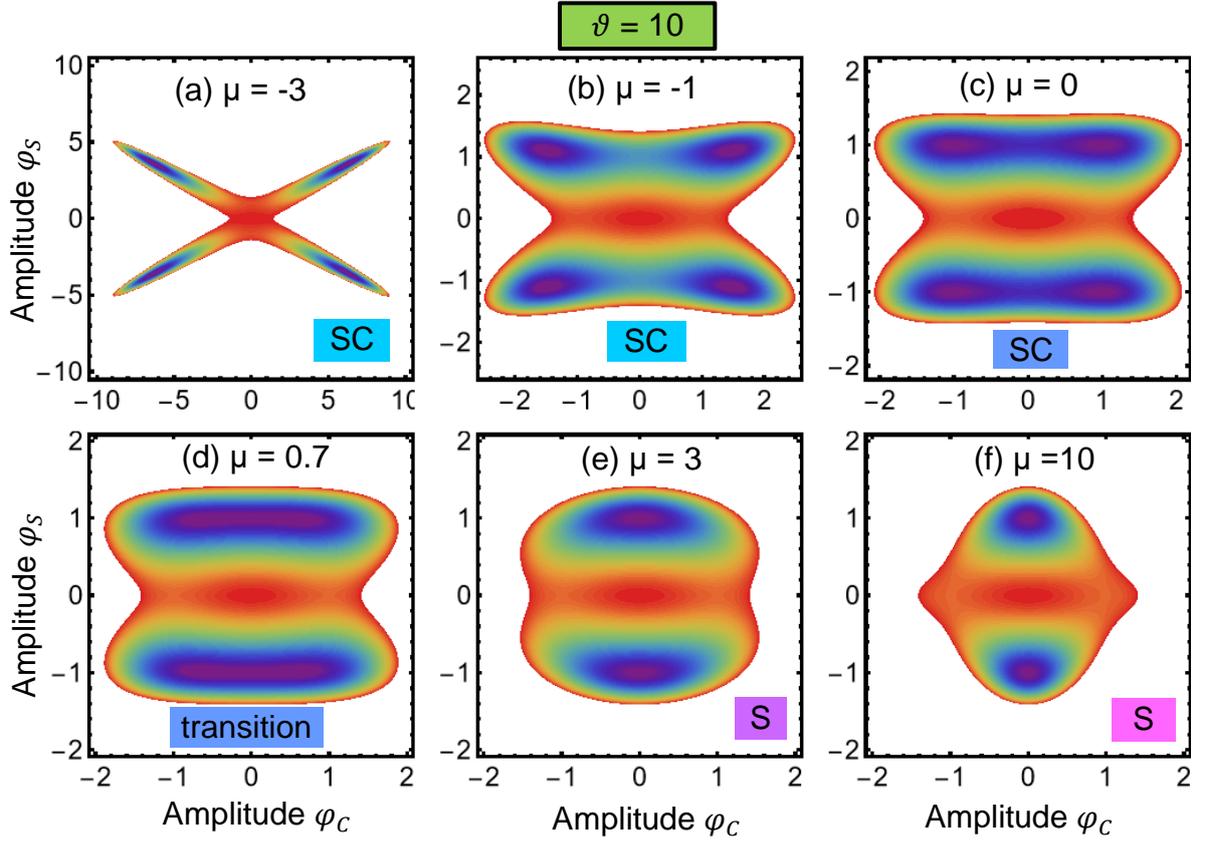

**FIGURE S2.** The free energy (S.7), as a function of order parameter amplitudes $\varphi_C$ and $\varphi_S$, calculated for different values of parameter $\mu$: **(a)** $\mu = -3$, **(b)** $\mu = -1$, **(c)** $\mu = 0$, **(d)** $\mu = 0.7$, **(e)** $\mu = 3$ and **(f)** $\mu = 10$. Red color denotes zero energy, while violet color is its minimal density in relative units. Capital letters "SC" and "S" denote the mixed spontaneous long-range order - charge order state, and spontaneous long-range ordered state, respectively. Other parameters: $\vartheta = 10$ and $\theta_C \approx \theta_S \approx -1$.

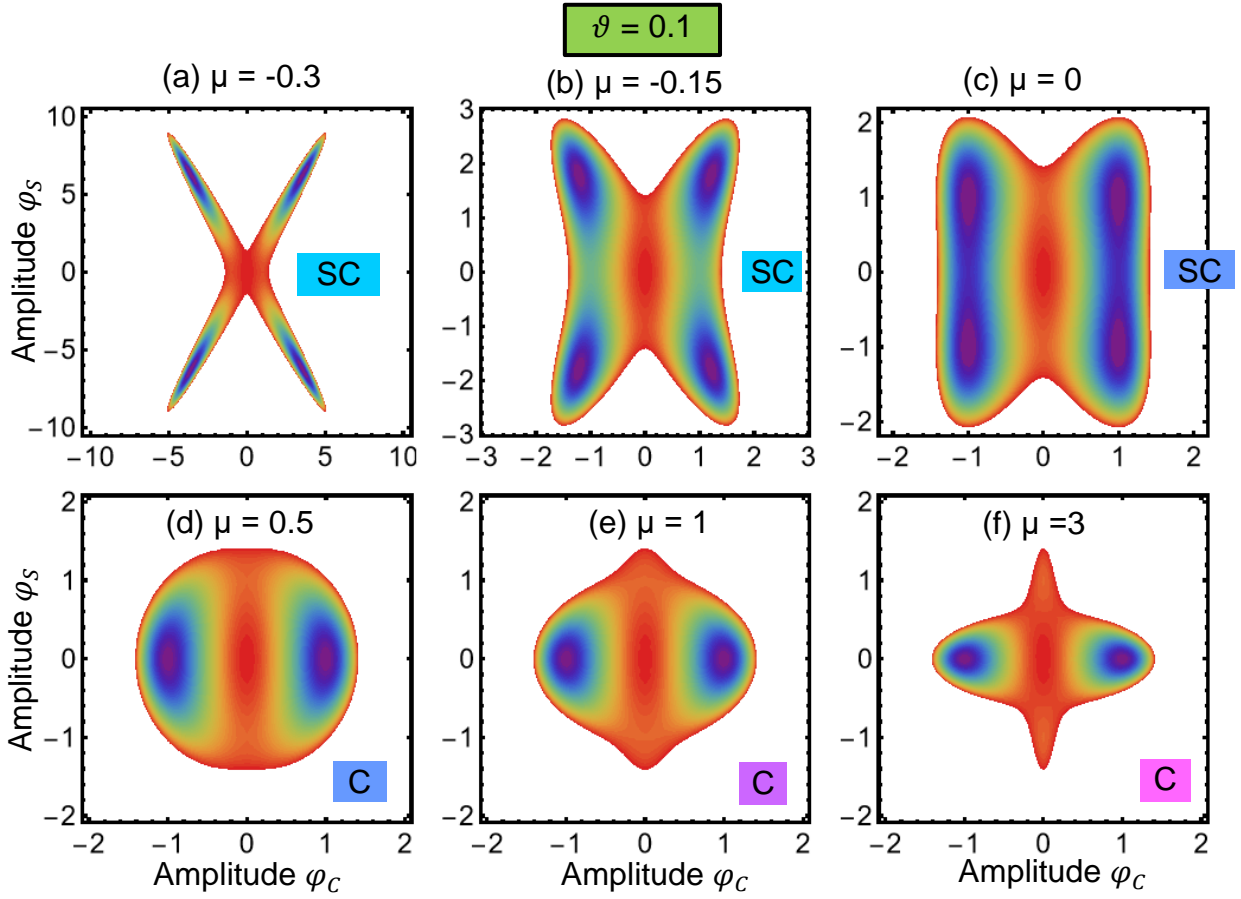

**FIGURE S3.** The free energy (S.7), as a function of order parameter amplitudes $\varphi_C$ and $\varphi_S$, calculated for different values of parameter $\mu$: **(a)** $\mu = -0.3$, **(b)** $\mu = -0.15$, **(c)** $\mu = 0$, **(d)** $\mu = 0.5$, **(e)** $\mu = 1$ and **(f)** $\mu = 3$. Red color denotes zero energy, while violet color is its minimal density in relative units Capital letters "SC" and "C" denote the mixed spontaneous long-range order - charge order state, and charge order state, respectively. Other parameters: $\vartheta = 0.1$ and $\theta_C \approx \theta_S \approx -1$.

### Appendix S3. Phase diagrams at zero temperature

Phase diagrams, as a function of $\eta^*$ and $\vartheta$, calculated from the free energy Eq.(S.7) for $T = 0$, $\chi^* = 1$, $v_c^* = v_s^* = 10$, $w_c^* = w_s^* = 0.1$, and several values of $\xi_s^* = \xi_c^* = \xi^*$ are shown in the top row of **Fig. S4**. Stable phases are absent in the white region, where the inequality $\eta^* > -\sqrt{\vartheta}$ is invalid, and so the free energy (S.7) is unstable. The regions of the modulated CW-S, SW-S, and SCW phases are largest for $\xi^* = -1$; they decrease monotonically with $\xi^*$ increase from -1 to -0.25. Instead, area of the homogeneous SC phase increases monotonically with $\xi^*$ increase from -1 to -0.25; and the phase occupies the regions occupied of the modulated at more negative $\xi^*$. The positions and areas of the homogeneous S and C phase regions are independent on $\xi^*$ value. The S phase is stable at $\vartheta > 1$, and the C phase is stable at $\vartheta < 1$. The boundary between the S and C phases is a horizonal line $\vartheta = 1$.

Dimensionless wavenumbers, $q_c$ and $k_s$, corresponding to the diagrams in **Fig. S4(a), (d)** and **(g)**, are shown in the middle and bottom rows of **Fig. S4**, as a function of $\eta^*$ and $\vartheta$. The wavenumber $q_c$ is nonzero in the SCW and CW-S phases, and $k_s$ is nonzero in the SCW and SW-C phases. The wavenumber $q_c$ tends to zero at the boundaries between CW-S and SC phases, and between SCW and SW-C phases; it continuously changes at the CW-S and SCW boundary, and increases strongly at the boundary with the white region, where the inequality $\eta^* > -\sqrt{\vartheta}$ is invalid, and so the free energy (S.7) is unstable. The wavenumber $k_s$ tends to zero at the border of SCW and CW-S phases; it continuously changes at the SW-C and SCW boundary, and increases strongly approaching the boundary between SW-C and C phases.

Note, that the phase diagrams in **Fig. S4(a), (d)** and **(g)** are constructed based on the equality of the free energies of corresponding phases, and only the regions of phases absolute stability are shown in different colors. On the wave vector diagrams, shown in **Fig. S4(b), (e)** and **(h)**, the vanishing of $q_c$ is shown, i.e., the region, where the SCW phase is metastable and the SW-C phase is absolutely stable is also shown, since the transition between the SCW and SW-C phases is a first-order transition. The coexistence region of the SCW and SW-C phases are marked by a semi-transparent background in **Fig. S4(b)**.

Normalized order parameters (or/and their amplitudes in the spatially-modulated phases), $\varphi_c$ and $\varphi_s$, are shown in the middle and bottom rows of **Fig. S5,** as a function of $\eta^*$ and $\vartheta$. As anticipated, the parameter $\varphi_c$ is zero in the S phase, and $\varphi_s$ is zero in the C phase. Both parameters are nonzero in the SCW, CW-S, SW-C, and SC phases, where their amplitudes are relatively small in comparison with the values near the boundary of SCW and SW-C phases with unusable white region, where the order parameters formally diverge.

Phase diagrams, as a function of $\xi^*$ and $\vartheta$, calculated from the free energy Eq.(S.7) for $T = 0$, $\chi^* = 1$, $v_c^* = v_s^* = v^* = 10$, $w_c^* = w_s^* = w^* = 0.1$, and several values of $\eta^*$ are shown in the top row of **Fig. S6**. Stable phases are absent in the white region, where the inequality $\xi^* > -\sqrt{\vartheta v^*}$ is invalid, and so the free energy (S.7) is unstable. The dark-green region of the SCW phase is biggest for $\eta^* = -0.5$; it decreases monotonically with $\eta^*$ increase from negative values to zero. At the same time, the small region of SW-C phase appears with $\eta^*$ increase from negative values to zero. The SC, CW-S and SW-C phases disappear at $\eta^* \geq 1$, being "adsorbed" by the regions of S and C phases. The S phase is stable at $\vartheta > 1$, and the C phase is stable at $\vartheta < 1$. The boundary between the S and C phases is a horizonal line $\vartheta = 1$. A very small triangular region of the SCW phase, located at the S-C boundary, maintains at $\eta^* = 1$.

Dimensionless wavenumbers, $q_c$ and $k_s$, corresponding to the diagrams in **Fig. S6(a), (d)** and **(g)**, are shown in the middle and bottom rows of **Fig. S6,** as a function of $\xi^*$ and $\vartheta$. The wavenumber $q_c$ is nonzero in the SCW and CW-S phases, and $k_s$ is nonzero in the SCW and SW-C phases. The wavenumber $q_c$ tends to zero at the boundaries between CW-S and SC phases [see **Fig. S6(b)**], and between SCW and SW-C phases [see **Fig. S6(e)**]; it continuously changes at the CW-S and SCW boundary, and increases approaching the boundaries of the SCW phase with the white region, where the inequality $\xi^* > -\sqrt{\vartheta v^*}$ is invalid, and so the free energy (S.7) is unstable. The wavenumber $k_s$ tends to zero at the border of SW-C and SCW phases with SC phase; it continuously changes at the SW-C and SCW boundary, and increases strongly approaching the boundary of the SW-C phase with the unstable region [see **Fig. S6(c, f, i)**].

Normalized order parameters (or/and their amplitudes in the spatially-modulated phases), $\varphi_c$ and $\varphi_s$, are shown in the middle and bottom rows of **Fig. S7,** as a function of $\xi^*$ and $\vartheta$. As anticipated, the parameter $\varphi_c$ is zero in the S phase, and $\varphi_s$ is zero in the C phase. Both parameters are nonzero in the SCW, CW-S, SW-C, and SC phases, where their amplitudes are relatively small in comparison with the values near the boundary of SCW and/or SW-C phases with unusable white region, where the order parameters $\varphi_c$ and/or $\varphi_s$ formally diverge.

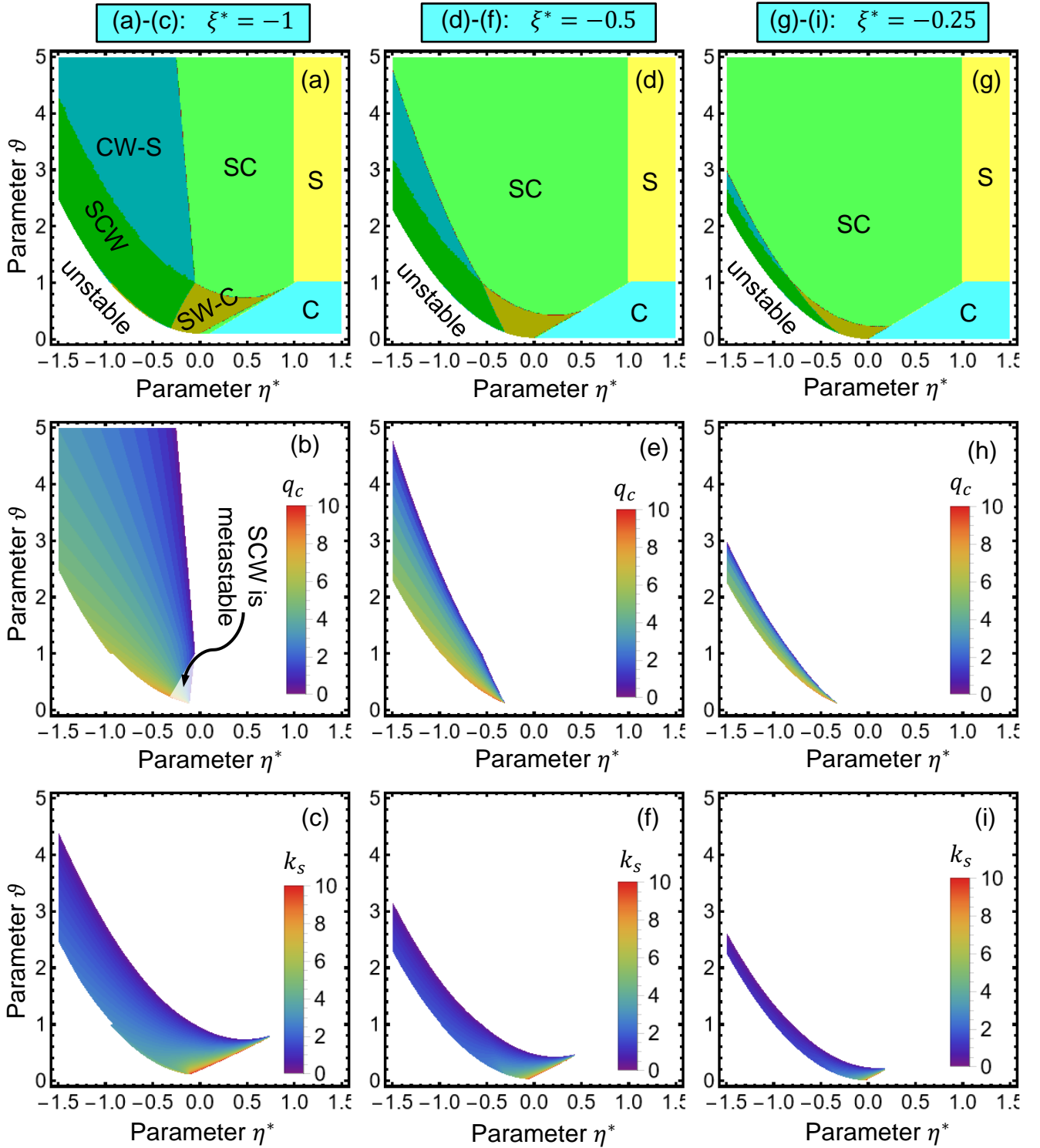

**FIGURE S4.** Phase diagrams (**a, d, g**), corresponding dimensionless wavenumbers, $q_c$ (**b, e, h**) and $k_s$ (**c, f, i**), as a function of $\eta^*$ and $\vartheta$, calculated from Eq.(S.7) at $T = 0$, $\chi^* = 1$, $\nu^* = 10$, $w^* = 0.1$, and several values of $\xi^* = -1$ for plots **(a)-(c)**, −0.5 for plots **(d)-(f)**, and −0.25 for plots **(g)-(i)**. Capital letters "SC", "S", "C" "CW", "SW", and "SCW" denote the regions of mixed spontaneous long-range order - charge order, long-range order, charge order, charge density waves, spontaneous long-range order waves, and intertwined long-range order - charge density waves, respectively. Red color in the contour plots of wavenumbers corresponds to their maximal values, dark-violet color is their minimal values, white color is for their absence.

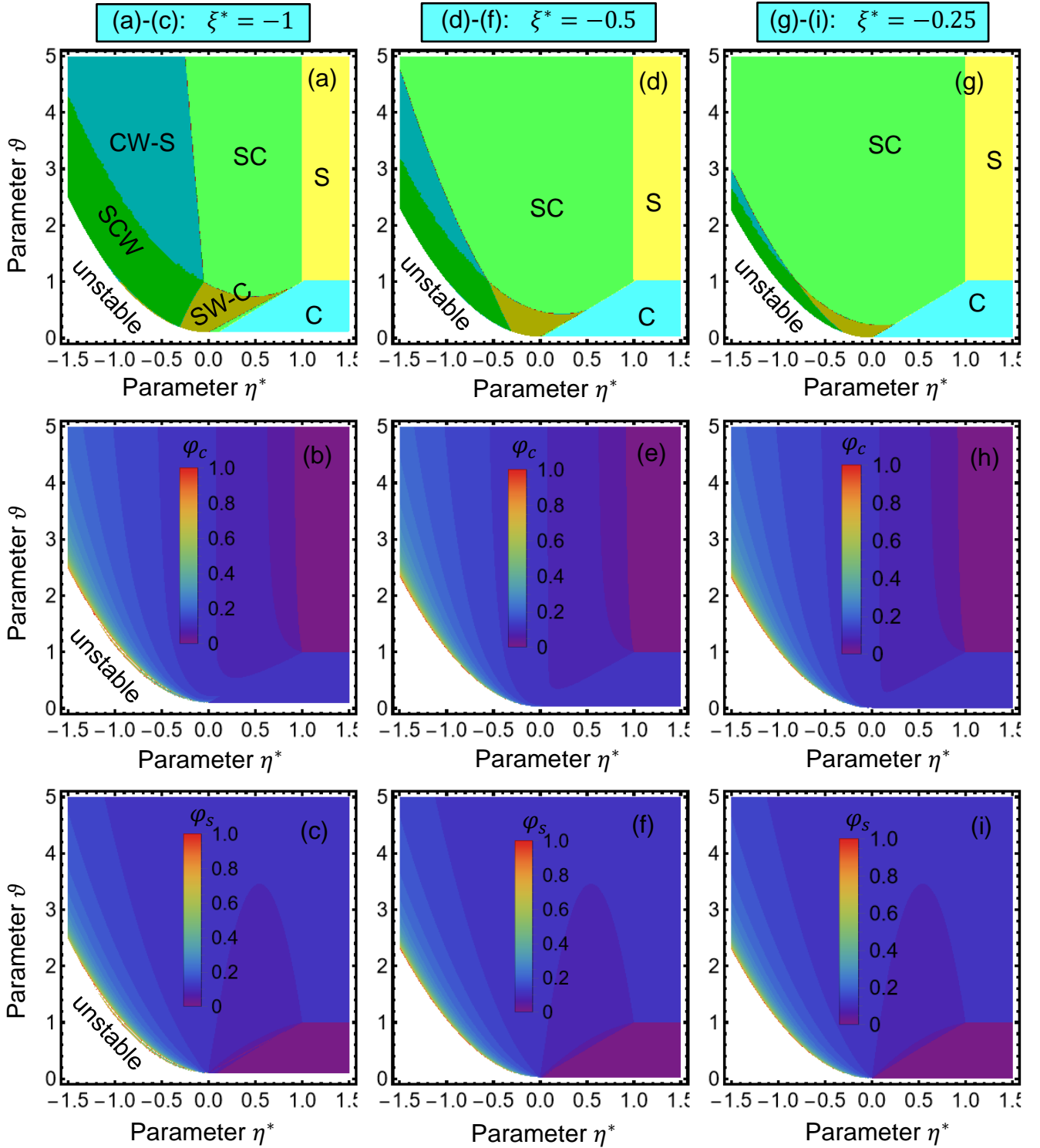

**FIGURE S5.** Phase diagrams (**a, d, g**), normalized order parameters (or/and their amplitudes), $\varphi_c$ (**b, e, h**) and $\varphi_s$ (**c, f, i**), as a function of $\eta^*$ and $\vartheta$, calculated from Eq.(S.7) at $T = 0$, $\chi^* = 1$, $v^* = 10$, $w^* = 0.1$, and several values of $\xi^* = -1$ for plots **(a)-(c)**, −0.5 for plots **(d)-(f)**, and −0.25 for plots **(g)-(i)**. Capital letters "SC", "S", "C" "CW", "SW", and "SCW" denote the regions of mixed spontaneous long-range order - charge order, long-range order, charge order, charge density waves, spontaneous long-range order waves, and intertwined long-range order - charge density waves, respectively. Red color in the contour plots of amplitudes corresponds to their maximal values, dark-violet color is their zero values.

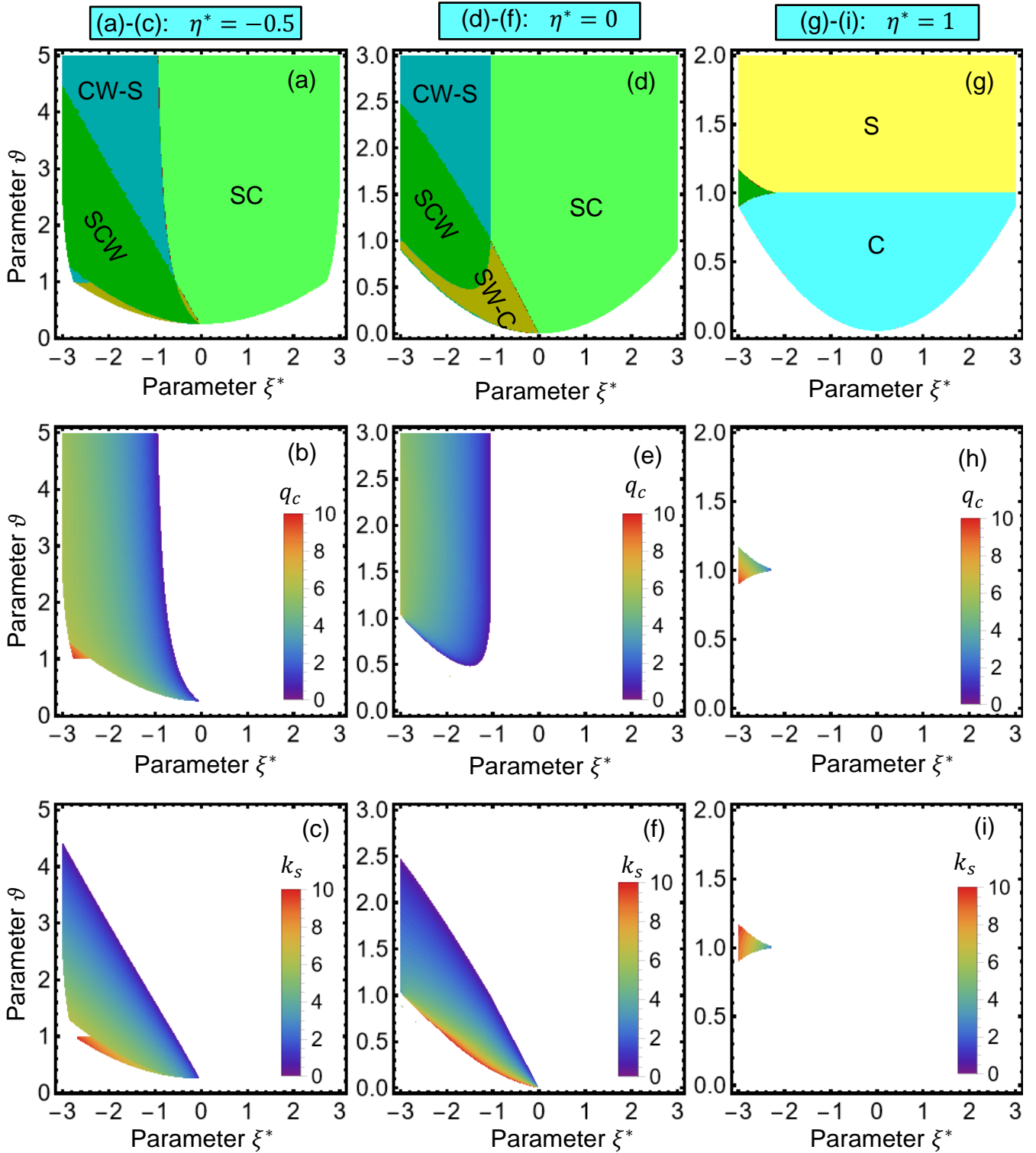

**FIGURE S6.** Phase diagrams (**a**, **d**, **g**), corresponding dimensionless wavenumbers, $q_c$ (**b**, **e**, **h**) and $k_s$ (**c**, **f**, **i**), as a function of $\xi^*$ and $\vartheta$, calculated the free energy Eq.(S.7) at $T = 0$, $\chi^* = 1$, $v^* = 10$, $w^* = 0.1$, and several values of $\eta^* = -0.5$ for plots **(a)-(c)**, $\eta^* = 0$ for plots **(d)-(f)**, and $\eta^* = 1$ for plots **(g)-(i)**. Capital letters "SC", "S", "C", "CW", "SW", and "SCW" denote the regions of mixed spontaneous long-range order - charge order, long-range order, charge

order, charge density waves, spontaneous long-range order waves, and intertwined long-range order - charge density waves, respectively. Red color in the contour plots of wavenumbers corresponds to their maximal values, dark-violet color is their minimal values, white color is for their absence.

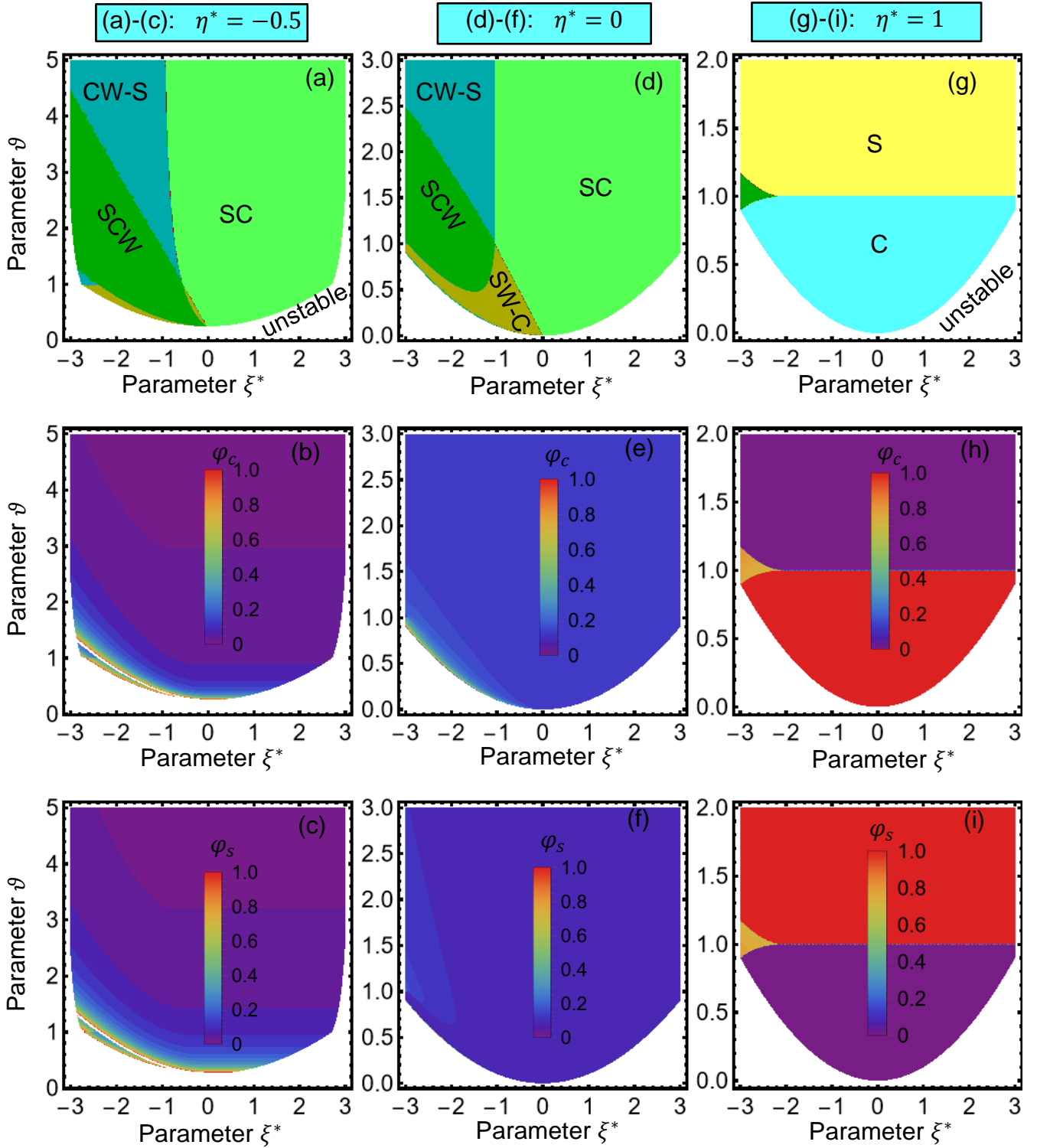

**FIGURE S7.** Phase diagrams (**a, d, g**), normalized order parameters (or/and their amplitudes), $\varphi_c$ (**b, e, h**) and $\varphi_s$ (**c, f, i**), as a function of $\xi^*$ and $\vartheta$, calculated the free energy Eq.(S.7) at $T = 0$, $\chi^* = 1$, $v^* = 10$, $w^* = 0.1$, and several values of $\eta^* = -0.5$ for plots (**a**)-(**c**), $\eta^* = 0$ for plots (**d**)-(**f**), and $\eta^* = 1$ for plots (**g**)-(**i**). Capital letters "SC", "S", "C", "CW", "SW", and "SCW" denote the regions of mixed spontaneous long-range order - charge order, long-range order, charge order, charge density waves, spontaneous long-range order waves, and intertwined long-range order - charge density

waves, respectively. Red color in the contour plots of amplitudes corresponds to their maximal values, dark-violet color is their zero values.